\documentclass[reprint,
 amsmath,amssymb,
aps,
]{revtex4-2}

\usepackage{caption}
\usepackage{subcaption}
\usepackage[utf8]{inputenc}

\usepackage{graphicx}
\usepackage{dcolumn}
\usepackage{bm}
\usepackage[breaklinks=true]{hyperref}
\usepackage{breakcites}
\usepackage[mathlines]{lineno}
\usepackage{float}
\usepackage[ruled,vlined]{algorithm2e}
\usepackage[english]{babel}

\usepackage{verbatim}

\usepackage[table]{xcolor}

\usepackage[autostyle]{csquotes}
\begin{document}

\preprint{APS/123-QED}

\title{Network Hierarchy and Pattern Recovery in Directed Sparse Hopfield Networks}

\author{Niall Rodgers}
\email{NXR081@student.bham.ac.uk}

 \affiliation{School of Mathematics, University of Birmingham, Birmingham B15 2TT, United Kingdom}

\affiliation{
 Topological Design Centre for Doctoral Training, University of Birmingham, Birmingham B15 2TT, United Kingdom}

\author{Peter Ti\v{n}o}

\affiliation{School of Computer Science, University of Birmingham, Birmingham B15 2TT, United Kingdom}

\author{Samuel Johnson}
 

\
\affiliation{School of Mathematics, University of Birmingham, Birmingham B15 2TT, United Kingdom and\\
 The Alan Turing Institute, British Library, 96 Euston Rd, London NW1 2DB, United Kingdom
}

\date{\today}

\begin{abstract}

Many real-world networks are directed, sparse and hierarchical, with a mixture of feed-forward and feedback connections with respect to the hierarchy.
Moreover, a small number of `master' nodes are often able to drive the whole system.
We study the dynamics of pattern presentation and recovery on sparse, directed, Hopfield-like neural networks using Trophic Analysis to characterise 
their hierarchical structure. This is a recent method which quantifies the local position of each node in a hierarchy (trophic level) as well as the global directionality of the network (trophic coherence). We show that even in a recurrent network, the state of the system can be controlled by a small subset of neurons which can be identified by their low trophic levels.
We also find that performance at the pattern recovery task can be significantly improved by tuning the 
trophic coherence and other topological properties of the network.
This may explain the relatively sparse and coherent structures observed in the animal brain, and provide insights for improving the architectures of artificial neural networks.
Moreover, we expect that the principles we demonstrate here, through numerical analysis, will be relevant for a broad class of system whose underlying network structure is directed and sparse, such as biological, social or financial networks.

\end{abstract}

\maketitle



\section{Introduction}

Models of the brain provided the original inspiration for the invention of artificial neural networks.
However, biological neural networks have a much richer structure than their artificial counterparts.  
In particular, they are not exclusively feed-forward like conventional deep network architectures, 
yet there is a direction to information processing,
unlike in
recurrent network models.
For example, the neural network of the nematode \textit{C. Elegans} \cite{White1986Elegans,Watts1998CollectiveNetworks} -- the only animal nervous system to have been fully mapped at the level of neurons and synapses -- is quite sparse and displays a non-trivial mix of 
feed-forward and feedback connections, as revealed by 
a recent technique from the field of complex networks called Trophic Analysis \cite{MacKay2020HowNetwork}. What might explain this particular neural-network architecture?  We address this question by studying
the relationship between trophic structure and the dynamics of a simple model which we refer to as a Hopfield-like neural network.


Trophic Analysis, inspired by ecological networks, assigns to each node a `trophic level', which can be regarded as a position in a hierarchy; and measures the `trophic coherence' of the whole network, a property which indicates to what extent this hierarchy is well defined, conferring to the network an overall directionality. In this work we take the convention that the bottom of the hierarchy is where information enters the system, just as energy flows up from plants in a food web. This may be different in other fields, for example in the study of `hierarchical trees', but all definitions are equivalent up to relabelling `top' and `bottom' or by reversing the edge directions.   
When the \textit{C.Elegans} neural network is visualised so as to show the trophic level of each neuron,
as in figure \ref{Worm_Net} in the appendix, it is observed that
while most of the synapses are consistent with an overall direction, there are some which feed back as in a recurrent architecture.
In fact, when the trophic coherence is calculated,
it lies exactly half way between a maximally coherent (i.e. entirely feed-forward) network, and one which is entirely incoherent (fully recurrent).
Moreover, it has been shown previously that this level of coherence amounts to a significant deviation from the kind of networks which arise from random graph models such as Erd\H{o}s–R\'{e}nyi model.  \cite{Erdos1959OnI,Johnson2017LooplessnessCoherence}.

How dynamics and hierarchy interact is demonstrated in this paper by performing a pattern recognition task (described in detail in the next section) on network architectures which span a range of hierarchical structures.
We find that trophic coherence is very strongly linked to the ability to correctly recognise and display the pattern shown. Maximally coherent networks lack the feedback to store patterns, while maximally incoherent networks are unable to change state when presented with fractions of new patterns. The optimal configuration is intermediate coherence, a mixture of feed-forward and feed-back structure which is shared with many biological systems. 
A similar result was reported in Ref. \cite{Johnson2020DigraphsSystems} for a system in which elements followed majority rule dynamics where the stability of the  \textit{C. Elegans} neural network was analysed using Trophic Analysis. We concentrate our study on synthetic networks inspired by this network which can be made dense enough to store multiple pattern states and lack basal nodes (nodes with no in-degree) which would act as input to the system without being influenced by it.


        	

There are clear differences between the structures of biological neural networks and artificial, recurrent neural networks, such as standard implementations of the Hopfield model. Biological networks are sparse, whereas the artificial versions are often based on complete or very dense graphs. They are also directed, since chemical synapses have a pre- and a post-synaptic neuron \cite{Kale2018EstimatingConnectomes}, while some models such as that of Hopfield tend to assume symmetric synapses in order to avoid the possible periodic or chaotic behaviour associated with asymmetric interactions \cite{Hopfield1982NeuralAbilities} or to align with experimental data limited to the undirected case \cite{Kale2018EstimatingConnectomes}.

And in nature there are a limited number of sensory neurons which receive information directly from the outside world, a fact not usually replicated in Hopfield models. 
However, it is possible to implement a Hopfield-like model on sparse, directed networks, and to present stimuli only to a subset of neurons, as we go on to do here in order to investigate how dynamics is affected by modifying the trophic structure.

Feed-forward artificial neural networks, such as those used in deep learning, in these respects resemble more closely the architectures of biological neural networks, at least in the case of nature's only fully mapped connectome, that of \textit{C. Elegans}.
 The main difference is that deep neural networks tend to be maximally coherent, with each layer corresponding to a distinct (integer) trophic level.

We show, through numerical analysis, that network hierarchy can be exploited in order to use a small subset of neurons to drive the system, with how well a pattern is recovered being strongly influenced by where in the hierarchy it is received. Hierarchical structure creates heterogeneous dynamics with different parts of the network recovering patterns differently.  Additionally, we show that by preferentially adding edges to lower level nodes, pattern recovery can be made more consistent. 
This has potential applications for how artificial neural networks are designed \cite{Zheng2010AnalysisDynamics, Cantini2019Hope4Genes:Data}, as well as for controllability of dynamics on general directed complex networks \cite{Szedlak2014ControlAttractors,Baggio2020EfficientNon-normality,Liu2016ControlSystems}, which could range from biological neural networks \cite{Lynn2019TheControl,Leonetti2020NetworkBrain}, to ecosystems, economies \cite{Klaise2016FromProcesses} or the Internet \cite{Chau2007IncorporatingSearching}. In particular, Hopfield networks have recently been used to model Gene Regulatory Networks \cite{Szedlak2017CellSystems,Rockne2020ModelingNetworks}.
We will therefore use this model to highlight principles which may be of general application to any system wired according to a directed network. This is the first work in which Trophic Analysis has been applied to Hopfield-like networks which have been trained to store patterns.

\section{Using Trophic Analysis to Quantify Network Hierarchy}

Trophic Analysis is a method of quantifying the hierarchy of nodes and the global directionality of a directed complex network, first introduced in 2014 \cite{Johnson2014TrophicStability}, which is based on the ecological concept of trophic level \cite{levine1980several}. A directed network, or graph, can be represented via an adjacency matrix, defined as:
 \begin{equation} \label{Adj_eq}
A_{ij}=
    \begin{cases}
     1  \text{ \quad if there exists an edge } i \to j \\ 
     0  \text{ \quad  otherwise }
    \end{cases}.
\end{equation}
Unlike in undirected networks, this matrix is not necessarily symmetric, $A_{ij}\neq A_{ji}$. Directed networks have the additional complexity of the notion of in- and out-degrees, where the in-degree is the number of incoming edges a vertex receives and the out-degree is the number of edges leaving a vertex. In  undirected networks the in- and out-degrees coincide.
Directed networks can also be weakly or strongly connected. Weakly connected means that there is a path between all pairs of vertices if you ignore the edge directions, while strongly connected means there is such a path respecting the edge directions. The networks studied in this work are all weakly connected but may not be strongly connected.

Trophic Analysis was recently extended and redefined to cover more general networks \cite{MacKay2020HowNetwork}, removing the requirement that networks must have basal nodes (nodes with in-degree 0). This is the definition that will be used in this work. Trophic structure has been used to study spreading processes in neural and epidemiological settings \cite{Klaise2016FromProcesses},  infrastructure  \cite{Pagani2019ResilienceNetworks,Pagani2020QuantifyingNetworks} and the structure of organisations \cite{Pilgrim2020OrganisationalAnarchy}. Trophic Analysis is composed of two parts: the node level information, trophic level,
which describes where each node sits in the overall hierarchy of a network; and the global information of how directed, or coherent, the overall network is. The idea of trophic level arises from ecology where the lowest trophic level nodes represent plants which sit at the bottom of the network hierarchy, and the highest trophic level nodes are carnivores at the top of the food chain. Trophic level can be calculated for a network of $N$ nodes by solving the $N \times N$ matrix equation \begin{equation} 
    \Lambda h = v,
    \label{eq_h}
\end{equation}
where $h$ is the vector of trophic levels, $v$ is the imbalance of in-degree and out-degree of a node, $ v_i = k_{i}^{in} - k_{i}^{out}$, and $\Lambda$ is the Laplacian matrix:
\begin{equation}
    \Lambda = diag(u) - A - A^{T}.
\end{equation}
This depends on the sum of the in- and out- degrees of each node, $ u_i = k^{in}_i + k^{out}_i$, the adjacency matrix, $A$, of the graph and its transpose, $A^{T}$. This definition can also be extended to cover weighted adjacency matrices \cite{MacKay2020HowNetwork}. The solutions to equation \ref{eq_h} can be modified by adding a constant vector since $\Lambda $ acting on a constant vector is zero. This allows the minimum level to be set at zero by convention and fully coherent networks to have integer levels.

Trophic coherence is based upon the distribution of trophic levels of the nodes in a network. How coherent or incoherent a network is can be described by the parameter \begin{equation}
    F = \frac{\sum_{ij}A_{ij}(h_{j} -h_{i}-1)^2}{\sum_{ij}A_{ij}}.
    \label{eq_F}
\end{equation}
We call $F$ the \textit{trophic incoherence}, such that when $F=0$ the network is completely coherent and when $F=1$ it is completely incoherent. This depends on the levels of each node $h_i$ and the entries of the adjacency matrix $A_{ij}$. 
Loosely speaking, $F$ quantifies, per connection in the graph, to what degree the connections $i \to j$ are not ``one-step'' connections in the order of trophic levels, i.e. by how much $(h_j-h_i)$ differs (in the mean square sense) from 1.
In principle these could have positive weights but throughout this work we will take the entries of the adjacency matrix to always be $0$ or $1$ to avoid confusion with the trained weights associated with the neural network. A network for which $F=0$ is acyclic and completely free from any feedback, with the amount of feedback and cycles growing as this parameter increases to 1 \cite{MacKay2020HowNetwork}. This is reflected in results showing an increase in spectral radius and a reduction in the deviation from normality of the adjacency matrix, how far the matrix is from commuting with its transpose, as incoherence increases \cite{Johnson2017LooplessnessCoherence,MacKay2020HowNetwork}. 

Note that the levels $h$, defined by Eq. (\ref{eq_h}), can be regarded as the argument which minimises $F$, as given by Eq. (\ref{eq_F}) \cite{MacKay2020HowNetwork}. One can therefore think of the trophic levels of a network as those which maximise its trophic coherence which relates to how it was derived in \cite{MacKay2020HowNetwork}.

\section{Hopfield-Like Networks}

The Hopfield Model is a recurrent neural network model which is very similar to the Ising model studied in statistical physics \cite{Hopfield1982NeuralAbilities}. The neurons can take binary states $+1$ or $-1$. Due to similarity to the Ising model these neuron states are sometimes referred to as spins and the order parameter measuring the state of the system can be referred to as a magnetisation.  A Hopfield network can store binary memories, or patterns, by setting the weights of connections between neurons such that when an update rule is applied the system moves across an energy landscape to its attractors, which correspond to the stored patterns.  This system can, in some cases, be studied via mean-field theory or other theoretical methods \cite{Amit1985Spin-glassNetworks}. In our case, however, due to the asymmetric connections and complex network topology,
we will use numerical simulations.  

We want the system to update in such a way that it moves towards the minima in the energy landscape 
defined by \begin{equation}
    E = - \sum_{ij} w_{ij}A_{ij}s_{i}s_{j},
\end{equation}
where $ w_{ij} $ is the coupling between neurons $i$ and $j$, which may be positive or negative depending on patterns stored. The states of the neurons take values $s_{i} = \pm 1$ and $A_{ij}$ are the elements of the adjacency matrix, as defined by Eq. (\ref{Adj_eq}).  There are many possible update rules which can achieve the desired behaviour, such as the Metropolis–Hastings algorithm \cite{Hastings1970MonteApplications}.
We use a sigmoid probability function such that \begin{equation} s_{i}(t+ \Delta t) = - s_{i}(t )\end{equation} with \begin{equation}
    \text{probability} = \frac{1}{1 + \exp{\frac{\Delta E}{T}}},
\end{equation}
where $\Delta E $ is the energy change associated with flipping the neuron state and $T$ is a temperature parameter which makes the system stochastic. To reduce complexity and uncertainty, the results we present here are for a temperature very close to zero, $T= 10^{-5}$, so the dynamics is essentially deterministic and equivalent to using the sign of the incoming field, the sum of the states of the in-neighbours, as the update rule.
The system can therefore be referred to as Hopfield-like, or simply as a Hopfield network, which is generally taken to be deterministic, as opposed to Boltzmann machines, which are stochastic \cite{Ackley1985AMachines}.
However, even in this regime the asymmetry in $A$ leads to a range of surprising behaviours not observed in undirected networks \cite{Crisanti1993TransitionModel}. 

 Updates to the system can be made in parallel or asynchronously. We use a parallel update rule, which allows for complex behaviour such as limit cycles \cite{Szedlak2014ControlAttractors}.

\subsection{Training the Network}

Setting the weights so that the attractors of the system correspond to the random binary patterns we wish to store in the network is a key part of the process. The traditional method of setting weights in a  Hopfield network so that the network recalls the desired patterns is Hebb's rule \cite{Attneave1950TheTheory}. This is often summarised as \enquote{neurons that fire together wire together}. That is, if two neurons have the same state in a particular pattern the connection between them is strengthened, and if they are in opposite states it is decreased. For learning $P$ patterns, where for each pattern each neuron has a fixed state $\xi^{p}_{i} = \pm 1$, the rule sets the weights as \begin{equation}
    w_{ij} = \frac{1}{P}\sum^{P}_{p=1} \xi^{p}_{i} \xi^{p}_{j}.
\end{equation} 
This very simple rule works and can be used on any network topology. It has the benefit of being a \enquote{one shot} rule in that it only requires one loop over the set of patterns to train the network. However, it suffers from the fact that on a graph which is not complete the information about the correlations between disconnected neurons is not used. We found during initial tests that on very sparse directed networks the memory capacity of the network was substantially reduced. This is very similar to the finding of Tanaka et al. \cite{Tanaka2020SpatiallyMemory} for undirected networks. They remedy this issue by adopting an iterative version of Hebb's rule based on earlier work \cite{Diederich1987LearningRules, Gardner1988TheModels} which was found to increase capacity substantially, with other similar results noted in the literature \cite{Davey2004HighConstraints}. For the remainder of this work we implement this rule \cite{Tanaka2020SpatiallyMemory}. Both the original Hebb rule and the adapted version are local, in that synaptic weights are updated using only information from the pre- and post-synaptic neurons -- as also happens, we believe, in the brain \cite{markram1997regulation}.

The iterative Hebb rule works to set the weights so that every pattern corresponds to a local minima of the energy landscape where updates of the system stop. 
This condition can be expressed as \begin{equation}
    \xi^{p}_{i}\left( \sum_{j} A_{ji} w_{ji}  \xi^{p}_{j} \right) \geq \delta 
\end{equation} for all $P$ patterns and $N$ nodes, and $\delta$ a positive constant. This means that at each node, for every pattern the polarities of the state and the incoming field are the same.
As a result it is always energetically unfavourable to flip the state at zero temperature so the system is stable.

The iterative Hebb rule is laid out in detail in Algorithm \ref{IterrativeHebbRule}.
\begin{algorithm}

\SetAlgoLined

Set the initial weights $w_{ij} =0 $ for all nodes $i,j$. \\
Set the stop condition flag, $\text{flag}=0$. \\
Set the step counter, $\text{steps} = 0$.

\While{$\text{flag} =0$ and $\text{steps} < \text{steps}_{\text{max}}$}{   flag=1 \; \For{ \text{p in range} P}{\For{ i in range N}{field = 0 \; \For {j in range N}{$ \text{field} \leftarrow \text{field} +   A_{ji} w_{ji}  \xi^{p}_{j} $ } \If{$\text{field}\times(\xi^{p}_{j}) < \delta $} {\For{q in range N}{$w_{qi} \leftarrow w_{qi} + \frac{A_{qi}\xi^{p}_{q}  \xi^{p}_{i}}{N}$  \; flag = 0 }}}}  $\text{steps} \leftarrow \text{steps} + 1 $}

\caption{Iterative Hebb Rule \cite{Tanaka2020SpatiallyMemory}}
\label{IterrativeHebbRule}

\end{algorithm}

At each iteration the weights are updated by \begin{equation}
    w_{ji} \leftarrow w_{ji} + \frac{A_{ji}\xi^{p}_{j}  \xi^{p}_{i}}{N},
\end{equation}
until the required condition is met. For this study $\delta$ was always set at 1, but other values can be used to change the stability of the patterns. If a stable solution of this set of inequalities exists it should always converge in a finite amount of time \cite{Diederich1987LearningRules}. However, a solution does not always exist for sparse, directed networks, so the algorithm needs to be terminated after a chosen maximum number of iterations. Here we use 400 iterations. The patterns can still be quite successfully stored and recovered if full convergence has not been achieved, as the number of weights continually updated is small after only a few iterations. 

Pattern recovery is measured with an order parameter, which we call magnetisation, and is defined for each pattern $p$ as the scalar product of the state of the system and the pattern: \begin{equation}
    m_{p} = \frac{1}{N}\sum_{i = 1}^N s_{i}\xi^{p}_i.
\end{equation} 
This is equivalent to the cosine of the angle between the state and the pattern. In this work we study patterns which are random, independent and identically distributed. Correlation between patterns and between patterns and the network topology may affect the performance of the network in a wide variety of ways depending on the topology, sparsity and nature of the correlation \cite{Tanaka2020SpatiallyMemory,Gutfreund1988NeuralPatterns}, so this may be a potential avenue for future work.

\section{Network Generation}\label{sec:Network_Generation}

To generate networks with a specified trophic coherence and fixed numbers of nodes and edges, we use a variant of the Generalised Preferential Preying Model (GPPM) from Refs. \cite{Klaise2017TheWebs,Klaise2016FromProcesses}, although the original work used a different definition of trophic level. 

We generate networks such that each node has in-degree at least 1. One reason for this is that if the network contains basal nodes (nodes with in-degree 0), one must choose whether their states $s$ should remain constant, take random values at each time step, or act as external inputs to the system.
Moreover, it is known that basal (or source) nodes can drive dynamics on directed networks in certain contexts \cite{Johnson2020DigraphsSystems,Wright2019TheDynamics}; but, to the best of our knowledge, we investigate here for the first time the importance of trophic level for dynamics on networks without basal nodes.

The detailed steps of the generative process are laid out in appendix \ref{app:Network_Generation}. In short, we randomly generate an initial configuration of $N$ nodes where each node has in-degree 1 and then calculate the initial trophic level, $\tilde{h}$, of this configuration. Then edges are added until the specific number is reached where the probability of connecting node i to j is   \begin{equation}
        P_{ij}= \exp{\left[ -  \frac{(\tilde{h}_{j} - \tilde{h}_{j} -1 )^2}{2T_{\text{Gen}}}\right]}.  
    \end{equation}
Afterwards, the updated trophic levels, $h$, are recalculated. With this method networks of any incoherence can be generated by varying the control parameter $T_{\text{Gen}}$, as demonstrated in Fig. \ref{fig:generation} in the appendix. 

\begin{figure}[H]

\centering
      		
            \includegraphics[width=1.0\linewidth]{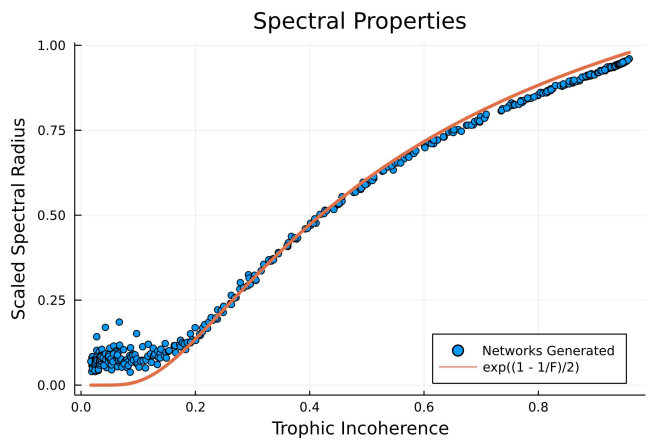}
        	\caption{Scaled Spectral Radius of Generated Networks against trophic incoherence following the same analytic prediction as real networks as shown in Ref. \cite{MacKay2020HowNetwork}. The number of nodes is always $N = 500$, and the mean degree $\langle k \rangle =20$. }
        	
        \label{fig:spectral_radius}
        \end{figure}

The networks generated via this method can act as an approximation to the hierarchical structures seen in real-world systems. In Ref. \cite{MacKay2020HowNetwork} it was shown that many real-world networks conform approximately to an analytical prediction for their scaled spectral radius, $\rho_{s}$, as a function of the incoherence parameter, $F$. This relationship is \begin{equation}
    \rho_{s} = \exp\left[{\frac{(1-\frac{1}{F})}{2}}\right],
    \label{eq_rho}
\end{equation}
and can be derived from the `coherence ensemble' of random graphs \cite{Johnson2017LooplessnessCoherence}.
Here, $\rho_{s}$ is defined such that it is scaled between 0 and 1 to compare networks of different sizes:
\begin{equation}
    \rho_s = \frac{\rho}{||A||_2},
\end{equation}
where $\rho$ is the standard spectral radius of the adjacency matrix, and $||A||_{2}$ is the 2-norm of $A$ -- that is, $||A||_{2}^2$ is the largest eigenvalue of $AA^T$.
As we show in Fig. \ref{fig:spectral_radius}, the generated networks we use in this work also have $\rho_s$ close to the value given by Eq. (\ref{eq_rho}). This justifies the assumption that the numerically generated networks reflect some of the characteristics exhibited by real world networks.

\section{Results}

\begin{figure}[H]

\centering

            \includegraphics[width=1.0\linewidth]{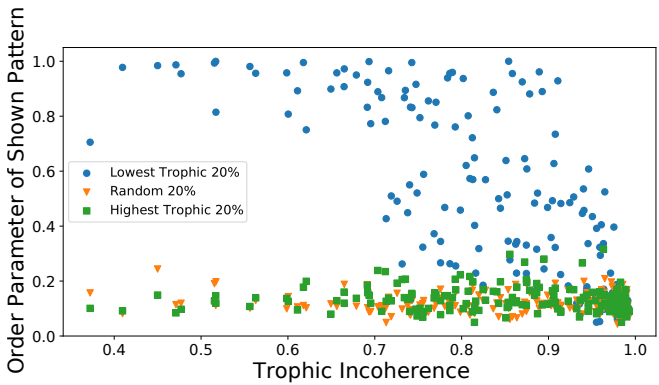}
            
        	\caption{Performance of 200 Networks with $N = 500$, $\langle k \rangle =100$ recovering 10 patterns plotted against trophic incoherence. Showing patterns to different 20\% sets of nodes. 
        	}
        	
        \label{fig:dense_networks_20}
        \end{figure}

Firstly, as we impose the constraint that only a subset of neurons are presented the pattern to make the setup more like real-world systems, we must decide which set of neurons are to be shown the pattern and assess the effects of this choice. It was chosen that for this model 20\% of the neurons would be set into a pattern state and then it would be measured how well the system recovered the remainder of the pattern from this setup. The location of pattern presentation is analogous to the initial conditions of a dynamical system where the question would be which initial condition sends the system into the desired state given the constraint of only controlling a small number of elements. To assess the effect of hierarchy on pattern recovery, patterns were shown to the 20\% of nodes with the lowest trophic level (at the bottom of the hierarchy), highest trophic level (at the top of the hierarchy), and a random 20\% of nodes. The results are shown in figure \ref{fig:dense_networks_20} for networks spanning a range of trophic coherence. We plot the results for each individual network, rather than just the averages with error bars, in order to highlight the breadth and distribution of network behaviour, which becomes more apparent as we study sparser networks in section \ref{sec:sparser_networks}. These results demonstrate the difference in dynamics depending on the part of the network shown the pattern. When the pattern is shown to 20\% of the nodes randomly this is not enough to move the system into a new state, so the shown pattern is not recovered well across the whole range of trophic coherence. It is only possible to extend down to networks of intermediate coherence at this edge density with the generative method used. When the perturbation is made to the state of the top 20\% of nodes by trophic level, it has little effect on the state of the system. This is because the perturbation cannot filter back down the system, so the top nodes do not drive the dynamics. For sparse enough networks and high coherence, it is unlikely there will be any paths from the highest trophic levels to other nodes further down. If the network is denser, such paths may exist, but they will still be few compared with the number of paths form lower levels to higher. Hence, information flow will always be predominantly form lower to higher trophic levels in coherent networks.

The dynamics are more complex when patterns are presented to the lowest level nodes, since we observe different behaviours when trophic incoherence is varied. For the most incoherent networks, which are most similar to random graphs, the performance is on average poorer as the system is more stable due to the amount of feedback in the system. By stability we mean here the system's resistance to changing state when a new pattern is presented. At intermediate coherence, the network has an overall direction, so the perturbation at low level nodes is transmitted through the hierarchical network structure and pattern recovery is quite good even though only 20\% of nodes are stimulated. This is behaviour that would not be seen in a Hopfield model on a complete graph, nor on a random graph, since more than half the nodes would need to be changed to a new pattern in order to change the state of the system. These results demonstrate the variety of dynamics that can be induced by the more complex, hierarchical networks as compared to a complete or random graph \cite{PerezCastillo2004TheGraph,Hopfield1982NeuralAbilities,Mceliece1987TheMemory}.

\begin{figure}[H]

            \includegraphics[width=1.0\linewidth]{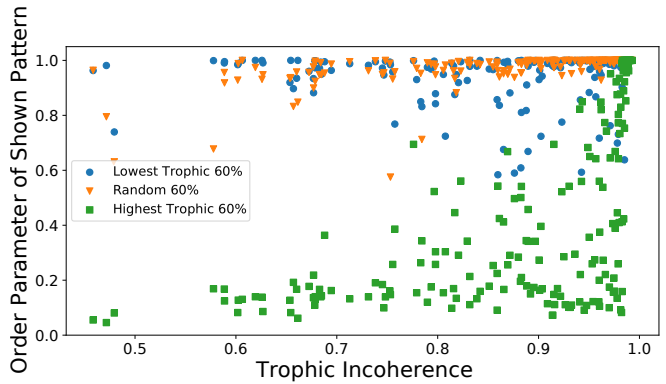}
        	\caption{Performance of 200 Networks with $N = 500$, $\langle k \rangle =100$ recovering 10 patterns plotted against trophic incoherence. Showing patterns to different 60\% sets of nodes.  
        	}
        	
        \label{fig:dense_networks_60}
        \end{figure}

When the constraint of a small number of input neurons is removed, the effect of hierarchy on the dynamics is less obvious. 
Figure \ref{fig:dense_networks_60} illustrates the case when the patterns are shown to 60\% of neurons.
When this many neurons receive an input the distinction between outcome of showing a pattern to a random 60\% and the lowest level 60\% is blurred, with both being able to recover the pattern across a range of trophic coherence. This highlights the impact of removing the constraint of a small number of inputs. When the inputs are large the effect of trophic level is hidden as randomly chosen inputs can control the system. However, for the highest level nodes this is still not the case. Even at 60\%, the higher level nodes fail to influence the coherent networks, as the lowest level nodes still have more control over the system and prevent the pattern from being modified. When the network is hierarchical, perturbations can be both amplified or damped by the structure, something we don't see in either a complete or a random graph Hopfield network. This is again different behaviour than what would be observed on a dense network with no internal structure, as 60\% of neurons being flipped would be enough to change the state to that of the new pattern in all cases. This highlights the connection between the trophic level of a node and its ability to control the network: the high level nodes have much less ability to influence the system than those at a low level. This difference remains at all levels of trophic coherence, but is most pronounced for more coherent structures. In all examples the trophic incoherence does not actually reach 1, where all the nodes would have the same level. This is because this only happens in balanced networks, such as a directed cycle, and the limit of our model is Erd\H{o}s–R\'{e}nyi random graphs, which have incoherence around 0.95. It is interesting to note the graphs which are random still have a slight hierarchical structure which can be revealed by the trophic levels.

\subsection{Sparser Networks} \label{sec:sparser_networks}

When the networks are made sparser -- that is, the average degree $\langle k \rangle$ is reduced from 100 to 20 -- the results are broadly the same as on denser networks, but there is more variation in the performance of different networks, even for similar trophic coherence. 
For networks of this sparsity the whole range of coherence can be investigated, as there are no difficulties associated with generating the more coherent networks. For inputs to both randomly selected and  highest level nodes, the recovery is very poor, just as it was before.
When it is the lowest level nodes which receive the input, behaviour depends on the trophic incoherence of the network. For the networks with lowest incoherence, the performance is generally very poor. This is due to the fact that these networks have very little feedback and small strongly connected components, so the patterns are not well recovered. For the intermediate coherence networks, performance is inconsistent. Some networks perform very well, with their structure being suited to controlling the system with only the low level nodes, while other networks perform very badly. Finally, higher incoherence networks are again more likely to get stuck in a pattern rather than to respond to the stimulus at the lowest level nodes, due to the high amount of feedback in the system, and the maximum performance begins to decrease again. Therefore, for sparser networks we find that the best performance is found at intermediate coherence -- although not all networks in this range are necessarily high performing.

\begin{figure}[h]

            \includegraphics[width=1.0\linewidth]{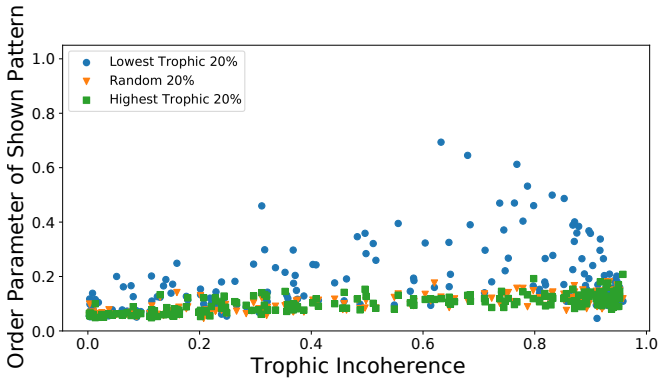}

        	\caption{Performance of 200 Networks with trophic incoherence showing patterns to the 20\% lowest (blue), highest (green) and random (orange) nodes by trophic level. $ N = 500$,  $ \langle k \rangle =20 $ }
        	
        \label{fig:unbiased_intial}
        \end{figure}

\begin{figure}[h]
      		
            \includegraphics[width=1.0\linewidth]{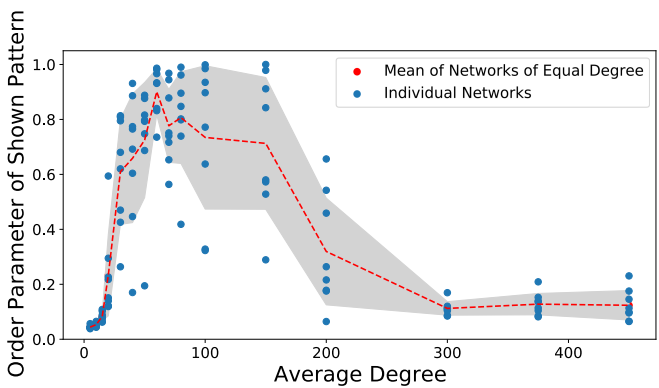}
        	\caption{Performance of Networks of varying degree for fixed generation temperature, $T_{\text{GEN}} = 1,$ showing patterns to the 20\% lowest  nodes by trophic level. $N = 500$. With the average trend shown by the dashed red line and the standard deviation shown in the shaded area.   
        	}
        	
        \label{fig:varying_degree}
        \end{figure}

The relationship between average degree and recovery of patterns is shown in figure \ref{fig:varying_degree}, where all networks have 500 nodes and are generated using $T_{GEN}=1$. The task cannot be performed by the most sparse networks, as they all fail to store any patterns. At an average degree of around 20, we reach the regime where some recovery is possible. For higher density, recovery reaches an inconsistent regime,  where performance varies greatly for networks of similar degree and trophic properties. This kind of regime is most interesting to study since the dynamics have a lot of variability, and successful pattern recovery is possible but not sure. Above an average degree of about 200, the structural features of the network are lost as the network is too dense and it simply gets stuck in one state for the whole dynamics and there is no ability to update when presented with a small number of inputs. Hence, figure \ref{fig:varying_degree} demonstrates that increasing the network density can make performance at a pattern recovery task worse, which is counter to the general expectation for Hopfield networks  where higher connectivity improves performance \cite{Sompolinsky1986NeuralNoise,Mceliece1987TheMemory}.

\subsection{Comparison of Targeting Highest Degree Nodes}  

\begin{figure}[h]

\begin{subfigure}{.5\textwidth}
      		
            \includegraphics[width=1.0\linewidth]{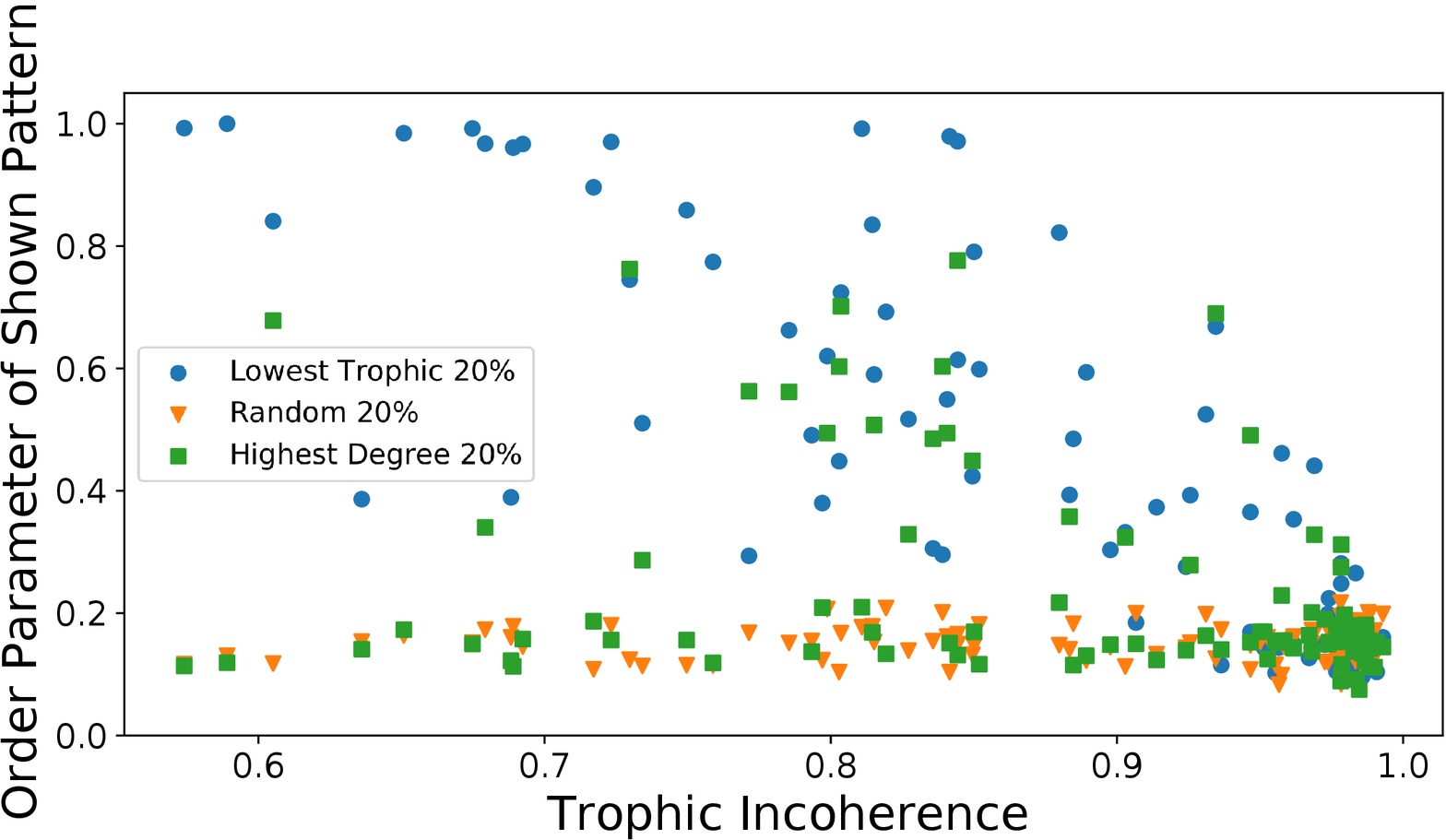}
        	\caption{
        	}
        	
        \label{fig:degree_high_dense}
        \end{subfigure}

\begin{subfigure}{.5\textwidth}
      		
            \includegraphics[width=1.0\linewidth]{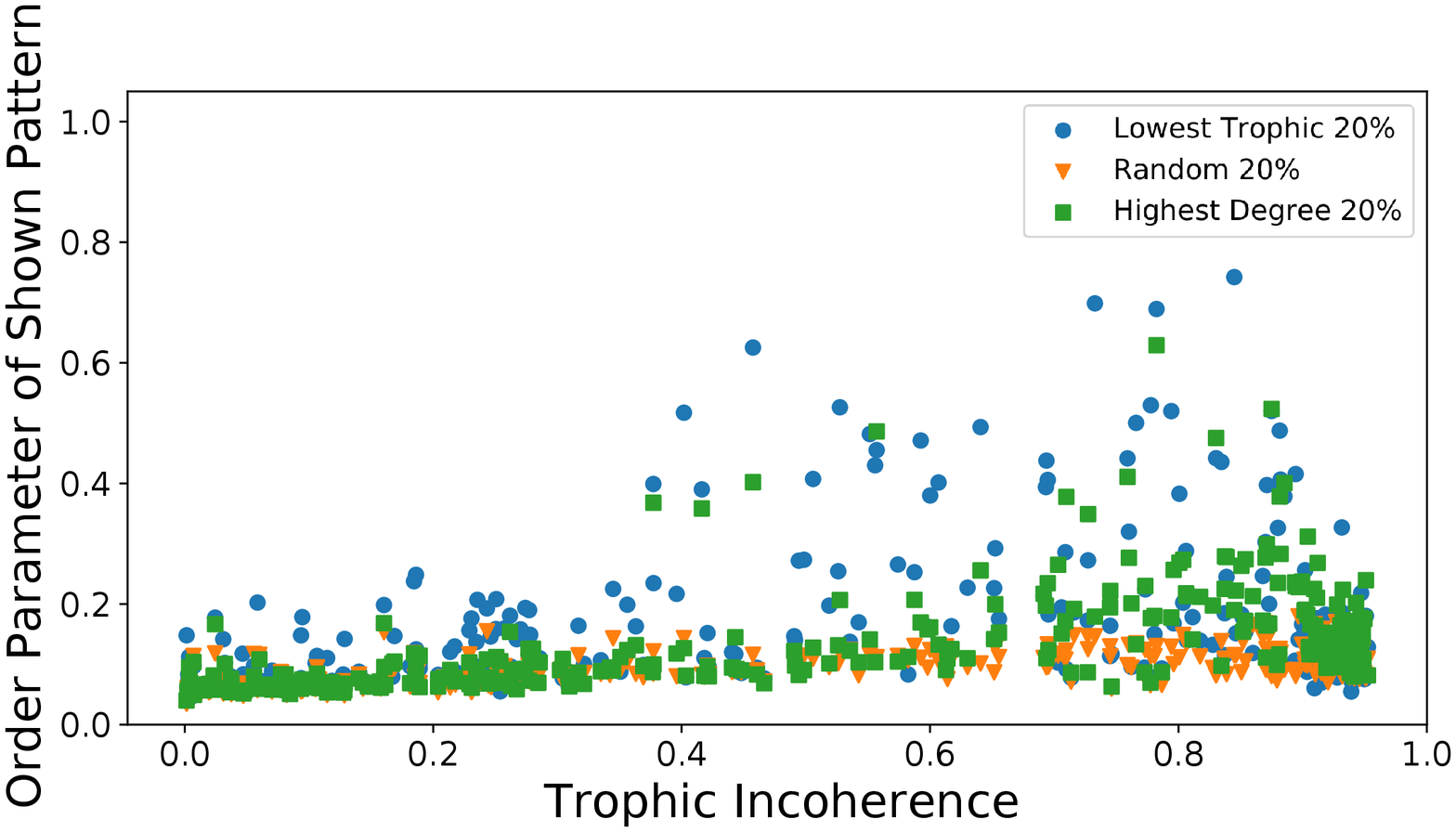}
        	\caption{
        	}
        	
        \label{fig:degree_high_sparse}
        \end{subfigure}

   \caption{Distribution of Performance for Networks showing 10 patterns to lowest trophic level, highest degree and random 20\% of nodes. N = 500 (a) $\langle k \rangle = 100 $, (b) $\langle k \rangle = 20$.}
    \label{fig:high_degree}
\end{figure}

To validate our choice of nodes we compare our results to a targeted presentation of the pattern to the 20\% of nodes of highest out-degree, which one might assume form the subset of nodes with greatest local influence on the system. This comparison is shown in figure \ref{fig:high_degree}, which compares the influence of the nodes of high degree to the selection of nodes by their local trophic properties. The set of nodes with highest degree do not influence the network to the extent that the lowest level nodes do. However, they do perform better than a random set of nodes, as expected, in both networks of average degree 20 and 100. In the networks of average degree 100 (figure \ref{fig:degree_high_dense}), the  lowest trophic level nodes are better than the highest degree nodes when the network is more coherent and hierarchical, as in this case the system is more strongly controlled by the low level nodes. When the networks are less hierarchical, the influence of the high out-degree nodes becomes comparable to the influence of the low level nodes. 
This highlights a crucial point: in a complex network, the ``importance" of nodes can be determined both 
by their degree-based centrality and by their relative position in the hierarchy, depending on how trophically coherent the overall system is.
In a very hierarchical (i.e. coherent) network, even if a node has a high out-degree, the state of the system can still be more controlled by lower out-degree nodes below it in the hierarchy. Our results, due to the generative model, focus on networks where the degree distributions are not extremely heterogeneous. The fact the in very hierarchical networks low level nodes control the state of the nodes above them would still hold in a very heterogeneous network. However, degree may be a more important factor if the out-degree of a few nodes were so large that they directly affected much of the network. These network properties can interplay in a variety of ways and may be the subject of future work.     

\subsection{Structural Properties of Networks Affecting Performance}

\begin{figure}[h]
      		
            \includegraphics[width=1.0\linewidth]{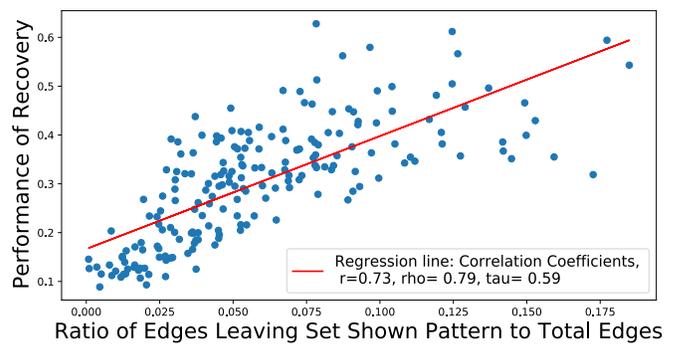}
        	\caption{Relationship between network performance and the ratio between number of edges leaving the set shown the pattern and total edges in the network. $ N = 500$, $\langle k \rangle = 20$. Networks of Intermediate Incoherence.    
        	}
        	
        \label{fig:edge_ratio}
        \end{figure}

We hypothesised that some network properties outwith trophic coherence could explain the range and inconsistency of behaviour for sparse networks. One possible measure was the number of edges leaving the node set shown the pattern compared to the total number of edges. When very few edges connect the nodes shown the pattern to the rest of the network, it is unlikely for the pattern to be successfully recovered, as when the pattern is updated it cannot be properly transmitted outside of the set shown the pattern. The results of this are displayed in figure \ref{fig:edge_ratio}. This shows that there is a strong correlation (correlation coefficients in the legend) between the edge ratio and performance, but it does not exactly determine the behaviour of the system. However, it is very clear the worse performing networks have very small values of this parameter, and it can be used to identify the failing networks, if not precisely to select the very best networks.

\begin{figure}[h]
      		
            \includegraphics[width=1.0\linewidth]{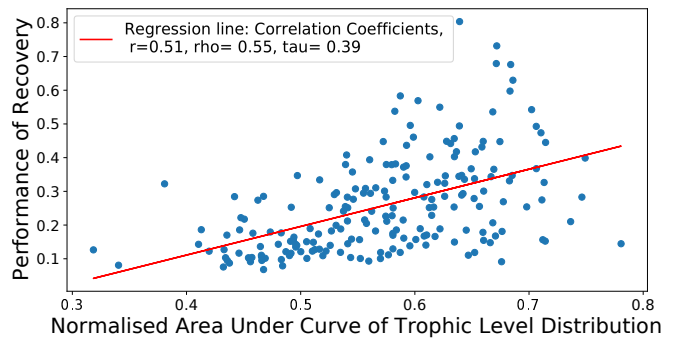}
        	\caption{Distribution of Performance of 200 Intermediate Incoherence Networks against the integral from 0 to 1 over a parameter $\alpha $ of the curve of number of nodes of trophic level less than $\alpha h_{max}$. $N = 500$, $\langle k \rangle = 20$ Networks of Intermediate Incoherence.    
        	}
        	
        \label{fig:alpha_sum}
        \end{figure}

Another factor which we thought may influence the  performance was the distribution of trophic levels amongst the nodes. In networks generated with the model used here (see section \ref{sec:Network_Generation}), edges tend only to span a small difference in trophic level. We would therefore like the level distribution to be peaked towards lower levels, so that more nodes have a lower level and are more likely to be densely interconnected with the set of nodes shown the pattern. This is shown in figure \ref{fig:alpha_sum}, where we sum the cumulative distribution of the number of nodes of trophic level less than $\alpha h_{max}$, for $\alpha $ in the range 0 to 1. This function is maximised when the level distribution peaks towards lower level nodes, and so provides a good measure of where the peak in trophic level lies, while being normalised so different networks can be compared. It shows a similar profile to the result of figure \ref{fig:edge_ratio}, where the correlation is again strong but does not precisely predict the performance of the network.  

We therefore surmise that the performance of a network at this task depends on several topological features, including but not limited to: trophic coherence, mean degree, mean degree of the lowest level nodes, and trophic level distribution.

\subsection{Time Series of Pattern Recovery in  Sparse Network Components}

In this section we review the time series of the dynamics of pattern recovery in a network with average degree 20, and highlight some of the reasons for the inconsistency in performance between similar networks. In all of the following example time series the pattern is presented to the 20\% of nodes with the lowest trophic level.

\begin{figure}[h]
  \centering    		
            \includegraphics[width=1.0\linewidth]{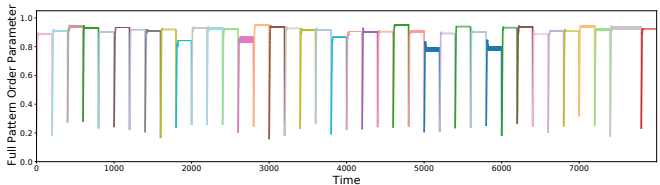}
        	\caption{  Time Series of Pattern Order Parameter for average degree 100 network with 500 nodes storing 20 patterns. Where each colour represents the pattern most recently having been shown to the network. At this edge density recovery is very consistent. }
        	
        \label{fig:dense_good_time_series}
        \end{figure}

We find that the network structure can induce quite heterogeneous dynamics. This is something that is not noticeable when the recovery is working well. 
Let us consider first the case of a dense network, with mean degree 100, as shown in figure \ref{fig:dense_good_time_series}. In this time series each colour represents the pattern which has been most recently presented to the network, while the y-axis represents the order parameter corresponding to that pattern. For a well performing network, the order parameter quickly returns near to 1 whenever a new pattern is presented. This is the case for the network shown in figure \ref{fig:dense_good_time_series}. Due to the recovery being this good, and the edge density being high, heterogeneous dynamics is not observed. Patterns are recovered to the same extent in all parts of the network hierarchy, and additionally the whole network is strongly connected, so there is no difference in dynamics inside or outside that component.

\begin{figure}
\begin{subfigure}{.5\textwidth}
    \includegraphics[width=1.0\linewidth]{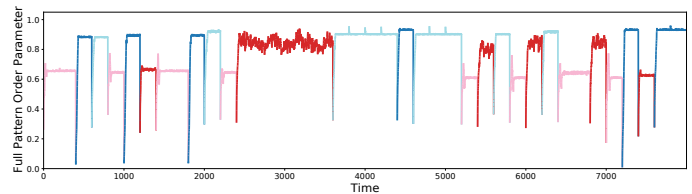}
        	\caption{   Full Pattern Order Parameter.  }
        	
        \label{fig:time_series_k_20_full}
\end{subfigure}

\begin{subfigure}{.5\textwidth}
 \centering    		
            \includegraphics[width=1.0\linewidth]{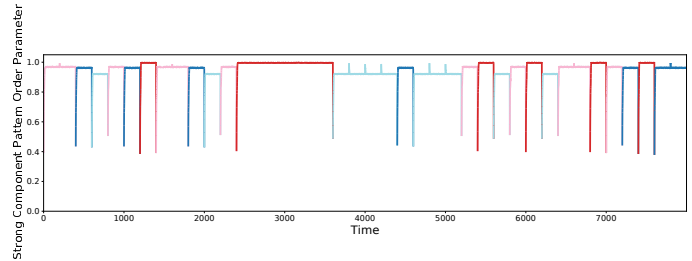}
        	\caption{ Strongly Connected Component.}
        	
        \label{fig:time_series_k_20_strong}
 
\end{subfigure}

\begin{subfigure}{.5\textwidth}
  \centering
  		
    \includegraphics[width=1.0\linewidth]{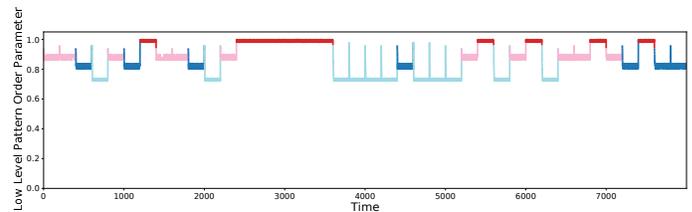}
    \caption{Bottom 20\% of nodes by trophic Level. }
        	
    \label{fig:time_series_k_20_low}

\end{subfigure}
\begin{subfigure}{.5\textwidth}
  \centering
  \includegraphics[width=1.0\linewidth]{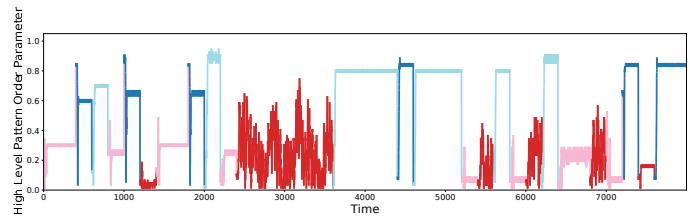}
        	\caption{  Top 20\% of nodes by trophic level.  }
        	
        \label{fig:time_series_k_20_high}
    \end{subfigure}
    
    \caption{Time Series for different network components for average degree 20 network with 500 nodes storing 4 patterns. Where each colour represents the pattern most recently having been shown to the network.}
    
    \label{fig:all_unbiased}
    
\end{figure}

This is very different from the dynamics exhibited by the sparse network used in figure \ref{fig:all_unbiased}, which stores four patterns. This network is a specific example of a network which performs reasonably well, but it should be borne in mind that many sparse networks fail very badly.  The dynamics are analysed by considering four different network components (subgraphs): the whole system; the largest strongly connected component; the bottom 20\% of the nodes by trophic level; and the top 20\% of nodes by level. Noting that these components are simply where the data was collected and the presentation location was unchanged. In the full system, figure \ref{fig:time_series_k_20_full}, recovery is good for some patterns but fails badly for others. The behaviour of each pattern is roughly consistent, and if a pattern fails or succeeds at one presentation it will repeat the same behaviour at subsequent presentations. The order parameter dips when a new pattern is presented, then moves to its new stable value. Additionally, there are fluctuations around the stationary states and updates to the system do not stop (i.e. some neurons continue to change state in subsequent time steps). This is different to the dynamics inside the largest strongly connected component, \ref{fig:time_series_k_20_strong}, where for the fully recovered stable patterns updates stop and there are no fluctuations. This highlights the stabilising effects of feedback associated with being strongly connected. Among the low level nodes, figure \ref{fig:time_series_k_20_low}, for those patterns which are correctly recalled, the order parameter goes to 1 when the new pattern is presented.
However, if a pattern is not recovered by the low level nodes then this precludes the possibility of that pattern being successfully transmitted through the network.
This means that if a pattern is not recovered by the low level nodes, figure \ref{fig:time_series_k_20_low}, then it will also fail to be recovered by the high level nodes \ref{fig:time_series_k_20_high}. Recovery by the high level nodes is the least consistent, and fluctuates the most, since these nodes are furthest from where the patterns are presented. In addition, the order parameter initially drops to zero whenever a new pattern is presented as it is not shown to any of the nodes contained in this set. 

These results might be different if the network included basal nodes (those with no in-neighbours), and would depend on what update rule we chose for these -- e.g. maintain their state indefinitely, update randomly, etc.

\subsection{Search for Improvements to Network Structure}

The results relating the distribution of trophic levels to  neural-network performance open the possibility of biasing the network generation process so that it preferentially leads to networks with topology better suited to the task. A simple way to accomplish this is to generate the networks in the same way as previously,  but modify the probability of adding edges so that it is biased towards adding edges to lower level nodes. This can be accomplished by modifying the probability of placing and edge so that \begin{equation}
    P_{ij}= \exp{\left[ -  \frac{(\tilde{h}_{j} - \tilde{h}_{j} -1 )^2}{2T_{\text{Gen}}} + \gamma \tilde{h}_{i}\right]},
\end{equation}
where the $\gamma \tilde{h}_{i}$ modification in the exponential acts to bias the distribution towards high or low levels, depending on the sign of $\gamma$. In what follows we choose $\gamma = - 0.5 $ in order to
add more edges to nodes with lower trophic level. One downside of this method is that it is harder to control precisely the trophic incoherence of a network and to span the full range of incoherence.

\begin{figure}[h]

\begin{subfigure}{.5\textwidth}
      		
            \includegraphics[width=1.0\linewidth]{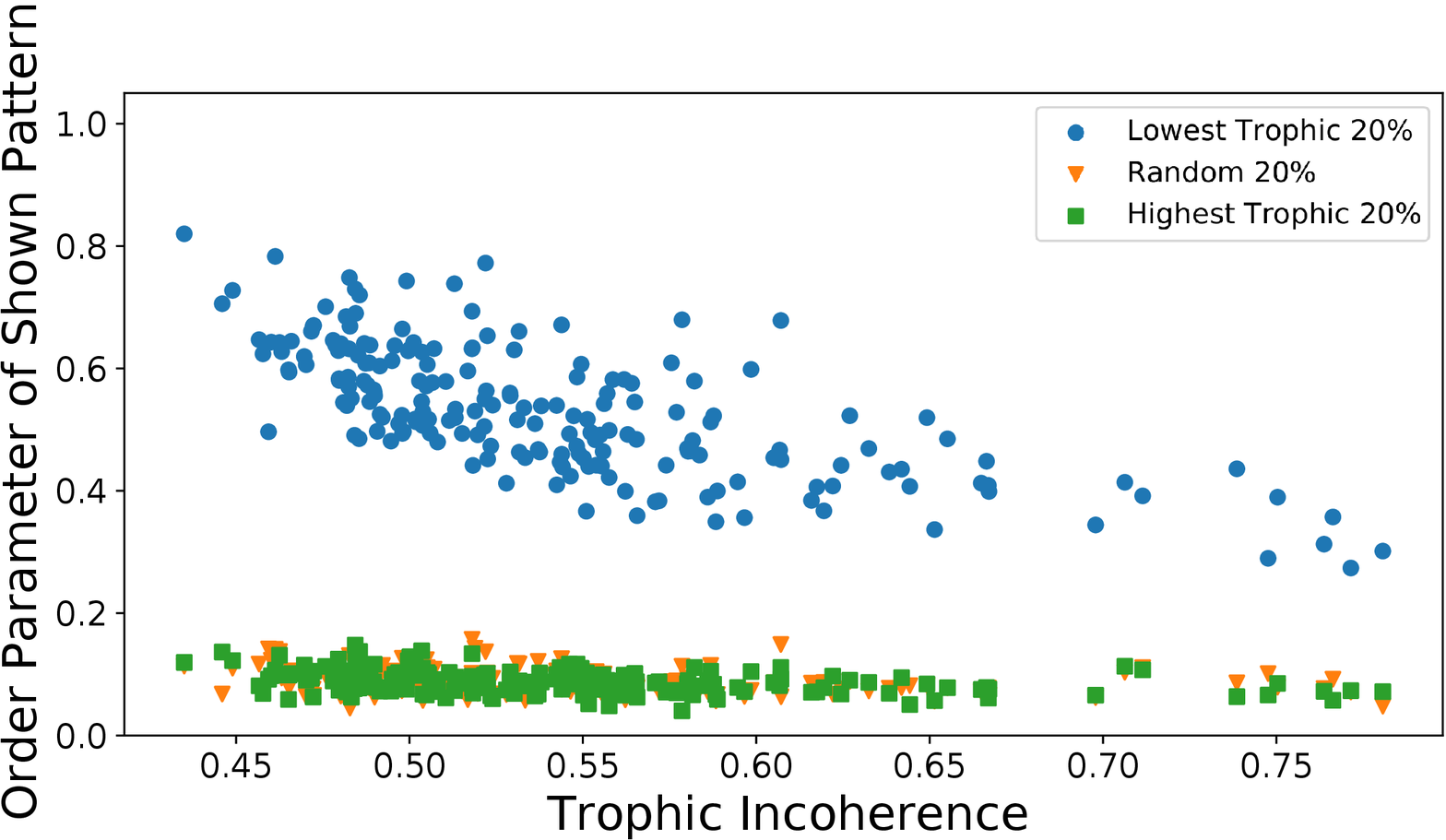}
        	\caption{Bias Towards Low Level Nodes. $\gamma = - 0.5 $
        	}
        	
        \label{fig:biased}
        \end{subfigure}

\begin{subfigure}{.5\textwidth}
      		
            \includegraphics[width=1.0\linewidth]{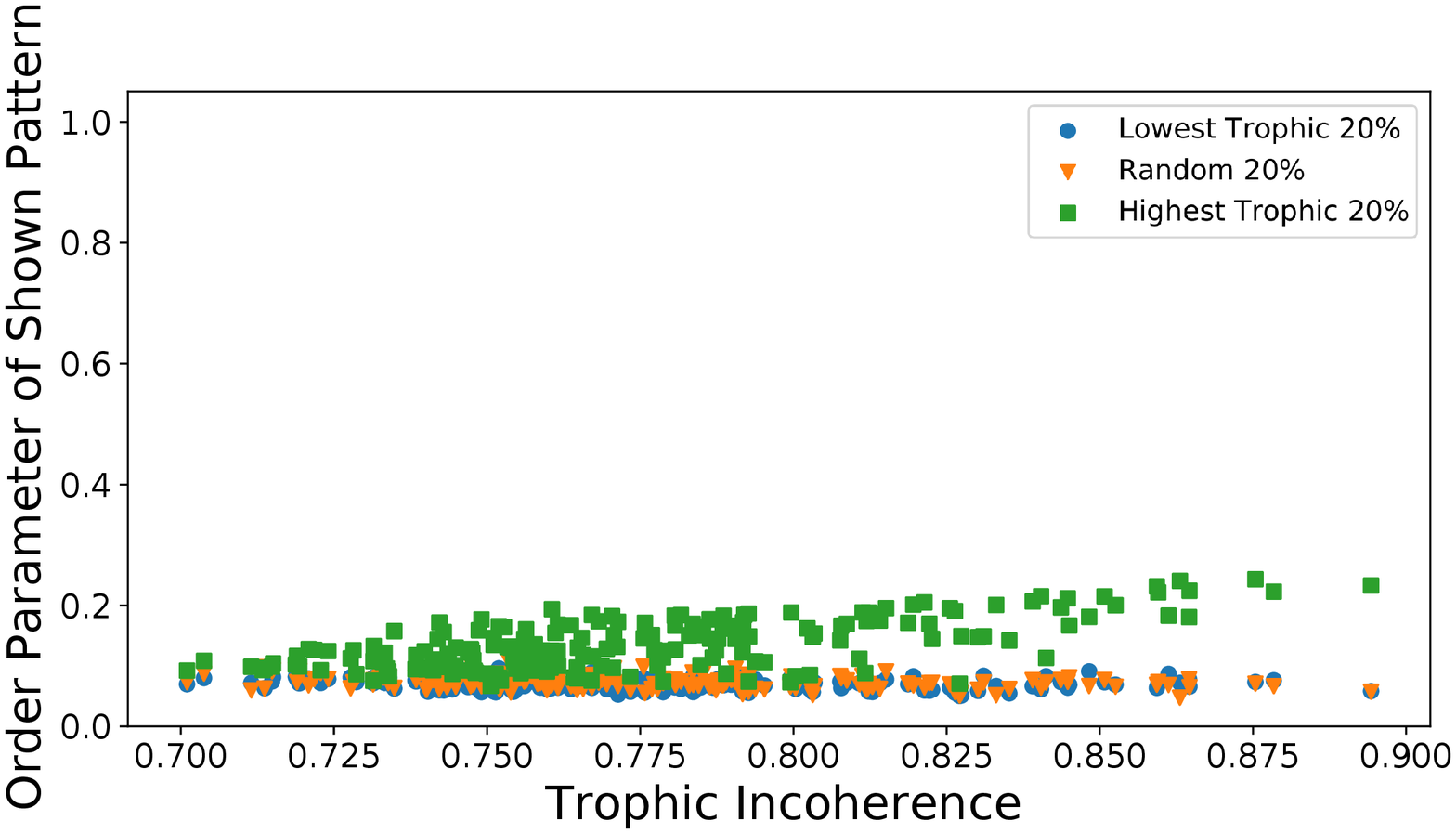}
        	\caption{Bias towards High Level Nodes.  $\gamma =  0.5 $
        	}
        	
        \label{fig:biased_high}
        \end{subfigure}

   \caption{Distribution of Performance for Biased Networks biased. N = 500 $\langle k \rangle = 20$.}
    \label{fig:biased_both}
\end{figure}

The broad effects of biasing the network generation and performance are demonstrated in figure \ref{fig:biased_both}. This shows that when edges are more likely to connect to low level nodes, figure \ref{fig:biased}, the very worse performing networks are essentially eliminated, and all the sparse networks recall at least a fraction of the pattern. This biasing has no effect on performance when presenting the pattern to the random or higher level nodes, as they still fail to force the system to change state when the perturbation is applied to these nodes.

To demonstrate the importance of where feedback is placed in the hierarchy we change the sign of the biasing factor and make it more likely that edges are added to the higher level nodes, figure \ref{fig:biased_high}. This creates networks which are not suited to the recovery task and perform badly in all cases. This is due to the fact that the edges connecting to high level nodes do not allow both for the pattern to be stable and for the information to be transmitted across the system. One issue with biasing the network generative process is that it becomes more difficult to control precisely the trophic coherence of the network, which is why the range of trophic incoherence is restricted in figure \ref{fig:biased_both}.

\begin{figure}[h]
\begin{subfigure}{.5\textwidth}
  \centering
  \includegraphics[width=1.0\linewidth]{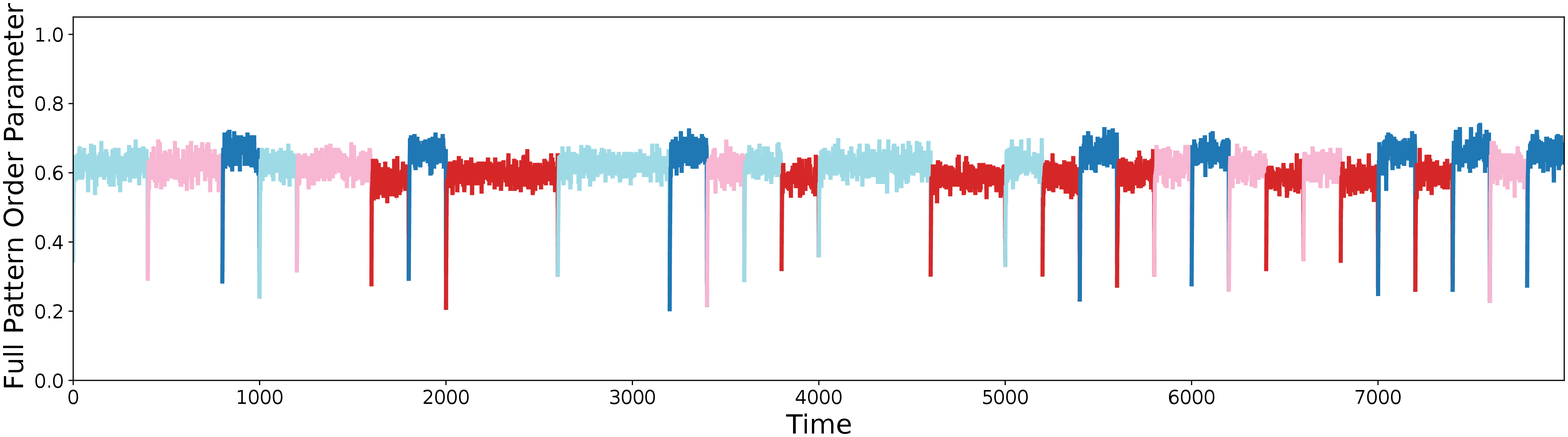}  
  \caption{Order Parameter of Whole Network}
  \label{fig:time_series_bias_full}
\end{subfigure}
\begin{subfigure}{.5\textwidth}
  \centering
  \includegraphics[width=1.0\linewidth]{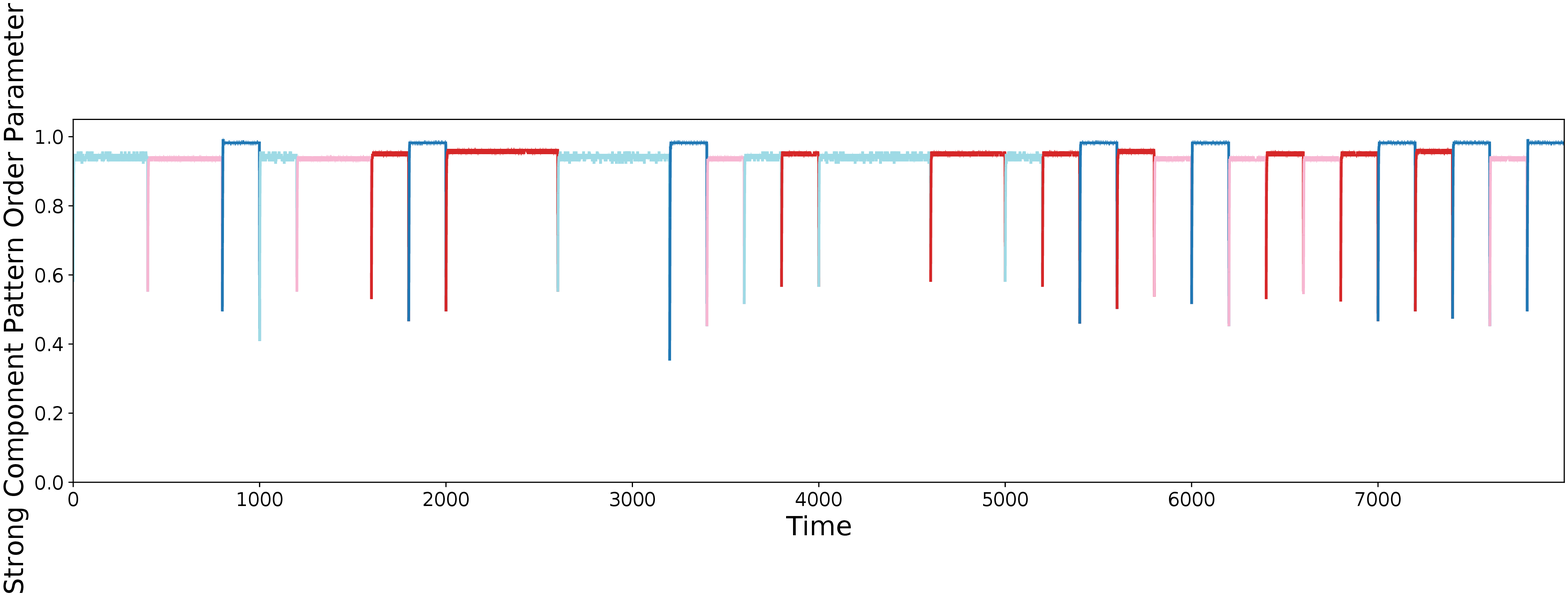}  
  \caption{Largest Strongly Connected Component Only}
  \label{fig:time_series_bias_strong}
\end{subfigure}
\caption{Time Series of Pattern Order Parameter calculated inside the different components for average degree 20 network with 500 nodes storing 4 patterns. Generated with a bias towards adding edges to lower level nodes.  Where each colour represents the pattern most recently having been shown to the network.}
\label{fig:time_series_biased}
\end{figure}

The time series of pattern recovery for sparse biased networks clearly demonstrate how this biasing procedure modifies the dynamics of the system. Pattern recovery across the whole system, figure \ref{fig:time_series_bias_full}, is very consistent compared to the unbiased networks  (figure \ref{fig:unbiased_intial}), which fully recover some patterns and fail to recover others. This time series is a representative example of the behaviour seen in biased networks and comes from a single network. The consistent level which they reach however is below 1 so the patterns are not fully recovered and the recovery is not as high as the maximum seen in some specific unbiased networks (figure \ref{fig:all_unbiased}). Whether this is better may depend on the context: remembering part of every pattern so it can be identified may be preferable to recalling some patterns perfectly but not recovering others at all. Additionally, for biased networks there are large fluctuations and updates continue when the system has reached the new state. This can be explained by looking at the dynamics inside the largest strongly connected component only, figure \ref{fig:time_series_bias_strong}. In this component recovery is very consistent and all patterns are fully recovered,
so the network does much better when this component is larger. It also explains why, in the time series for the dynamics of the full network, fluctuations around a stable point are observed, since updates to neuron states stop in the strongly connected component but continue outside of it. The fact that the recovery is very good inside the largest strongly connected component opens up the possibility of selectively generating networks which are both biased towards lower level nodes, and have large strongly connected components.

\begin{figure}[h]
      		
            \includegraphics[width=1.0\linewidth]{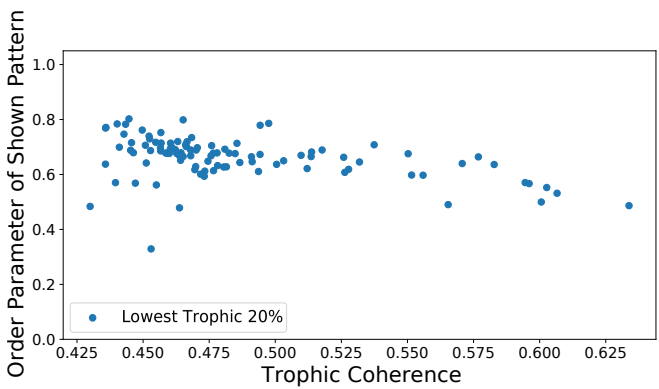}
        	\caption{Performance of Low Level Biased Networks with average degree 20 and 500 nodes where the largest Strongly Connected Component contains at least 60\% of the neurons.  
        	}
        	
        \label{fig:selective_generation}
        \end{figure}

This is demonstrated in figure \ref{fig:selective_generation}, which shows the performance of biased networks where the largest strongly connected component comprises more than 60\% of the nodes. These networks are simply generated by repeating the generative process and discarding networks which do not meet this requirement. The higher this threshold, the more inefficient the process but the more likely we are to keep only highly performing networks.
At at threshold of 60\% all very poorly performing networks are eliminated, and the recovery performance is consistently around 0.6.

These results demonstrate that despite the variability in the dynamics of directed sparse Hopfield networks, it is possible to
generate structures which perform well consistently 
by tuning a few parameters: $T_{GEN}$ to set the trophic coherence, $\gamma$ to place edges preferentially at lower level nodes, and the threshold for the minimum size of the strongly connected component.

There are many possible ways to modify network structure to maintain performance and we just give a small sample here. Biasing is limited by the fact it reduces the control of the trophic structure and that strongly biasing may decrease the size of the strongly connect component which is needed for recovery. The best way to improve network structure depends on the constraints, on whether edges can be added or removed, and on how success is defined. Biasing makes the recovery more consistent, but the performance of unbiased networks may peak higher for certain networks and patterns, which may be preferred in some situations.

\section{Discussion and Conclusion}

We have shown that neural networks based on 
sparse, trophically coherent graphs have a much wider 
range of possible behaviour
than ones based on either fully random or complete graphs \cite{PerezCastillo2004TheGraph,Hopfield1982NeuralAbilities,Mceliece1987TheMemory}, where all nodes necessarily have very similar dynamical roles.
This symmetry is broken in a coherent network, as different nodes can have very different abilities to affect the dynamics of the system. The interplay between
trophic structure
and dynamics has already been observed across a range of systems in the literature \cite{Johnson2014TrophicStability,Klaise2016FromProcesses,Pilgrim2020OrganisationalAnarchy}. 
It has also been shown that the coherence of a network is linked to the non-normality of the adjacency matrix \cite{Johnson2020DigraphsSystems,MacKay2020HowNetwork}.
Non-normality in networks has in turn been linked to sensitivity to perturbations and to the stability of the system across a wide range of dynamics \cite{Asllani2018StructureNetworks,Asllani2018TopologicalSystems,Muolo2019PatternsSystems,Muolo2020SynchronizationOptimality,Fruchart2021Non-reciprocalTransitions}, which is consistent with our results that more coherent networks are more sensitive to targeted perturbations and less stable.


The behaviour observed in the system studied here relies on two key facts: that the networks are sparse and that the sets of input nodes are small. If the networks are too dense then hierarchical structure is destroyed and the asymmetry between nodes does not exist (there is a limit to how coherent a dense network can be).
Moreover, it is thanks to the network's trophic coherence that a small subset of nodes is able to drive the dynamics of the whole system. 
Many real-world systems display both of these properties.
Additionally, they are often neither highly coherent nor incoherent, but have trophic coherence in the intermediate range which allows for a balance between stability and sensitivity to stimuli \cite{MacKay2020HowNetwork,Johnson2020DigraphsSystems,Johnson2017LooplessnessCoherence}.
Therefore, we believe the principles studied here for the case of Hopfield-like neural networks may be broadly applicably to a range of real-world systems. The limitations of these methods are that since trophic incoherence is an average global network property it lacks the precise detail to characterise fully the behaviour of the system in all cases. It is challenging to control precisely both the trophic incoherence and another aspect of network structure, since one may restrict the other, as with the biasing method. In future this work could be extended by looking at a time series of patterns which are correlated with each other \cite{Gutfreund1988NeuralPatterns}, patterns correlated with the structure, or networks with heterogeneous degree distributions and varying in- and out-degree correlations.  


In conclusion, we have demonstrated, through numerical analysis, that trophic structure strongly shapes pattern recovery in directed Hopfield-like networks. In particular, on a sparse, directed network a small number of input neurons -- which can be identified by their trophic levels even in the absence of basal nodes -- are able to drive the system in such a way that it recovers patterns. This would not be possible on either a complete or fully random network, which require at least about 50\% of the nodes to receive the input in order to change state.
In order for such networks to recover patterns successfully, they must have the correct balance between feedback and directionality -- a feature which is determined by the trophic coherence.
However, we observed that setting the appropriate trophic coherence was not enough to guarantee good performance.
We found that by biasing the network generation process so as to add edges preferentially to lower-level nodes, and then discarding networks with strongly connected components below a minimum size, we could reliably produce architectures that performed the task well.

\section*{Acknowledgment}

The authors would like to thank the Centre for Doctoral Training in Topological Design and Engineering and Physical Sciences Research Council (EPSRC) for funding this research. SJ also acknowledges support from the Alan Turing Institute under EPSRC Grant EP/N510129/1.

\appendix

\section{Software Tools Used}

 Various software packages were used manipulate the networks and perform the simulations. The Python package Graph Tool \cite{peixoto_graph-tool_2014} was used for some of the network manipulation. Networkx  \cite{Hagberg2008ExploringNetworkX} was used for Network drawing and some network manipulation and analysis. The Julia package LightGraphs.jl was used for the spectral radius results \cite{Bromberger17}. All the updating and training of the Hopfield-like networks was done with the aid of the Cython package \cite{Behnel2011Cython:Worlds} to convert Python Code to C as pure Python was found to be too slow to allow efficient study.

\section{Network Generation}\label{app:Network_Generation}

\begin{figure}[h]
      		
            \includegraphics[width=0.5\textwidth]{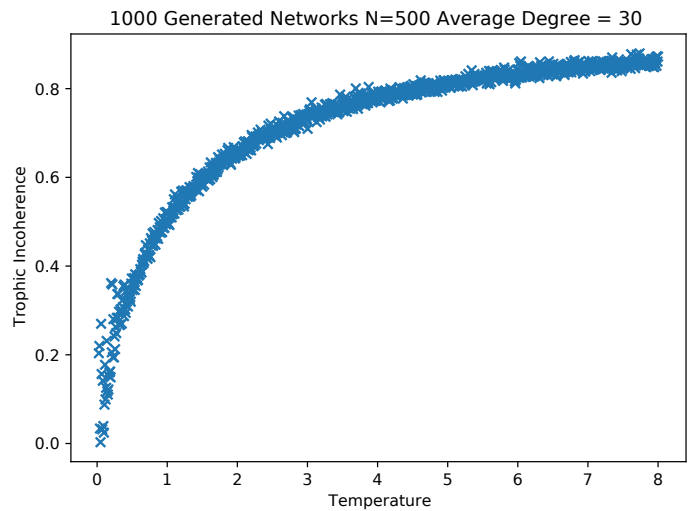}
        	\caption{An example of the distribution of trophic incoherence with temperature like parameter, $T_{\text{GEN}}$ in this generative model.
        	}
        	
        \label{fig:generation}
        \end{figure}

The detailed steps to the network generative process are as follows:

\begin{enumerate}
    \item Create the $N$ nodes of the network and assign to each node one in-coming edge, in each case from a randomly chosen other node. After this,
    each node has in-degree 1.

    \item Compute the initial trophic levels, $\tilde{h}$, using equation \ref{eq_h}. This is best solved iteratively,
    since this method is fast and works even is there are small disconnected components.
    
    \item Add edges up to the desired edge number with probability dependent on the trophic level difference between the nodes minus 1.
    The edge probability used was Gaussian and defined as \begin{equation}
        P_{ij}= \exp{\left[ -  \frac{(\tilde{h}_{j} - \tilde{h}_{j} -1 )^2}{2T_{\text{GEN}}}\right]},  
    \end{equation}   
    where $T_{\text{GEN}}$ is a temperature-like parameter used to control how coherent the network is: small $T_{\text{GEN}}$ generates networks which are highly coherent. 
    
    \item Recompute the trophic Levels, $h$, including the newly added edges. Then compute the incoherence parameter, $F$,  of the generated network.  
\end{enumerate}

This method works best for reasonably sparse networks, since when the edge density is too large it
becomes difficult to find configurations of high trophic coherence, if they exist at all.
On the other hand, if the edge density is very low the resulting network may not be even weakly connected. However, for a large range of  densities it will encounter no issues. Due to the stochastic nature of the method it is
is not possible to 
predict precisely the incoherence of a generated graph . For example 1000 networks generated with $500$ nodes and $30 \times 500$ edges, and temperature $T_{GEN}=1.3$, cluster around $F\approx 0.59$, with most networks in the interval $F\in  (0.56,0.65)$. However this level of precision is sufficient for analysing general regions of behaviour with no issues.

The third step can be quite computationally inefficient for large networks with many possible edges as the probabilities for adding an edge at most locations are very close to zero. This can be improved by more efficiently sampling the probability distribution using the method outlined below.

The goal of this sampling method is to set up the sampling so that each time a random number is drawn it results in an edge. This avoids repeatedly drawing numbers for the majority of edges which are unlikely to be added. The steps are:
\begin{enumerate}\itemsep=0em
    \item Label all the possible edges and probabilities with an integer $l$ and $P_{l}$ respectively.
    \item Compute the sum of all these probabilities, $$S = \sum_{l} P_{l}.$$ 
    \item Draw a random number, $r$, between 0 and S. 
    \item Sum the probabilities one at a time until you reach the random number, $r$.
    \item Add an edge at the space, $l$, corresponding to the probability $P_{l}$ which made the same greater than $r$.
    \item Set $P_{l} =0 $ and repeat sets 2-6 until the required edge number is reached. 
\end{enumerate}

This method is much more efficient: the sums can be computed quickly as it avoids the many repeated random number draws for every single missed edge that would otherwise be necessary.
It is possible to also create variants of this method by modifying the initial structure to which subsequent
edges are added; or to recast the model so as to start from a dense network and prune edges with a similarly defined probability to generate networks of the desired trophic incoherence.

\section{Connectome of \textit{C.Elegans} plotted by trophic level}

\begin{figure}[H]
      		\centering
            \includegraphics[width=0.5\linewidth, angle =270 ]{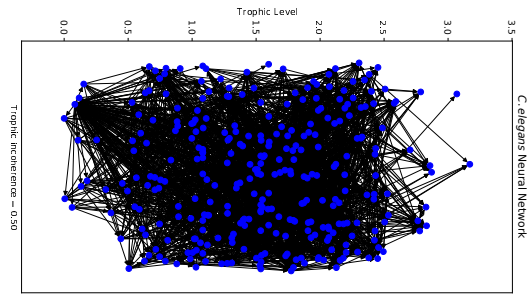}
        	\caption{Illustration of the the real world connectome of \textit{C.Elegans} shown which has intermediate incoherence with node height drawn using Trophic Level. Data from\cite{DataSamJohnson}. Drawn with Networkx Graph Package \cite{Hagberg2008ExploringNetworkX}.
        	}
        	
        \label{Worm_Net}
        \end{figure}

\bibliography{ref1,biblio}

\begin{thebibliography}{50}%
\makeatletter
\providecommand \@ifxundefined [1]{%
 \@ifx{#1\undefined}
}%
\providecommand \@ifnum [1]{%
 \ifnum #1\expandafter \@firstoftwo
 \else \expandafter \@secondoftwo
 \fi
}%
\providecommand \@ifx [1]{%
 \ifx #1\expandafter \@firstoftwo
 \else \expandafter \@secondoftwo
 \fi
}%
\providecommand \natexlab [1]{#1}%
\providecommand \enquote  [1]{``#1''}%
\providecommand \bibnamefont  [1]{#1}%
\providecommand \bibfnamefont [1]{#1}%
\providecommand \citenamefont [1]{#1}%
\providecommand \href@noop [0]{\@secondoftwo}%
\providecommand \href [0]{\begingroup \@sanitize@url \@href}%
\providecommand \@href[1]{\@@startlink{#1}\@@href}%
\providecommand \@@href[1]{\endgroup#1\@@endlink}%
\providecommand \@sanitize@url [0]{\catcode `\\12\catcode `\$12\catcode
  `\&12\catcode `\#12\catcode `\^12\catcode `\_12\catcode `\%12\relax}%
\providecommand \@@startlink[1]{}%
\providecommand \@@endlink[0]{}%
\providecommand \url  [0]{\begingroup\@sanitize@url \@url }%
\providecommand \@url [1]{\endgroup\@href {#1}{\urlprefix }}%
\providecommand \urlprefix  [0]{URL }%
\providecommand \Eprint [0]{\href }%
\providecommand \doibase [0]{https://doi.org/}%
\providecommand \selectlanguage [0]{\@gobble}%
\providecommand \bibinfo  [0]{\@secondoftwo}%
\providecommand \bibfield  [0]{\@secondoftwo}%
\providecommand \translation [1]{[#1]}%
\providecommand \BibitemOpen [0]{}%
\providecommand \bibitemStop [0]{}%
\providecommand \bibitemNoStop [0]{.\EOS\space}%
\providecommand \EOS [0]{\spacefactor3000\relax}%
\providecommand \BibitemShut  [1]{\csname bibitem#1\endcsname}%
\let\auto@bib@innerbib\@empty
\bibitem [{\citenamefont {White}\ \emph {et~al.}(1986)\citenamefont {White},
  \citenamefont {Southgate}, \citenamefont {Thomson},\ and\ \citenamefont
  {Brenner}}]{White1986Elegans}%
  \BibitemOpen
  \bibfield  {author} {\bibinfo {author} {\bibfnamefont {J.~G.}\ \bibnamefont
  {White}}, \bibinfo {author} {\bibfnamefont {E.}~\bibnamefont {Southgate}},
  \bibinfo {author} {\bibfnamefont {J.~N.}\ \bibnamefont {Thomson}},\ and\
  \bibinfo {author} {\bibfnamefont {S.}~\bibnamefont {Brenner}},\ }\bibfield
  {title} {\bibinfo {title} {{ The structure of the nervous system of the
  nematode Caenorhabditis elegans }},\ }\bibfield  {journal} {\bibinfo
  {journal} {Philosophical Transactions of the Royal Society of London. B,
  Biological Sciences}\ }\textbf {\bibinfo {volume} {314}},\ \href
  {https://doi.org/10.1098/rstb.1986.0056} {10.1098/rstb.1986.0056} (\bibinfo
  {year} {1986})\BibitemShut {NoStop}%
\bibitem [{\citenamefont {Watts}\ and\ \citenamefont
  {Strogatz}(1998)}]{Watts1998CollectiveNetworks}%
  \BibitemOpen
  \bibfield  {author} {\bibinfo {author} {\bibfnamefont {D.~J.}\ \bibnamefont
  {Watts}}\ and\ \bibinfo {author} {\bibfnamefont {S.~H.}\ \bibnamefont
  {Strogatz}},\ }\bibfield  {title} {\bibinfo {title} {{Collective dynamics of
  "small-world" networks}},\ }\bibfield  {journal} {\bibinfo  {journal}
  {Nature}\ }\textbf {\bibinfo {volume} {393}},\ \href
  {https://doi.org/10.1038/30918} {10.1038/30918} (\bibinfo {year}
  {1998})\BibitemShut {NoStop}%
\bibitem [{\citenamefont {MacKay}\ \emph {et~al.}(2020)\citenamefont {MacKay},
  \citenamefont {Johnson},\ and\ \citenamefont
  {Sansom}}]{MacKay2020HowNetwork}%
  \BibitemOpen
  \bibfield  {author} {\bibinfo {author} {\bibfnamefont {R.~S.}\ \bibnamefont
  {MacKay}}, \bibinfo {author} {\bibfnamefont {S.}~\bibnamefont {Johnson}},\
  and\ \bibinfo {author} {\bibfnamefont {B.}~\bibnamefont {Sansom}},\
  }\bibfield  {title} {\bibinfo {title} {{How directed is a directed
  network?}},\ }\href {https://doi.org/10.1098/rsos.201138} {\bibfield
  {journal} {\bibinfo  {journal} {Royal Society Open Science}\ }\textbf
  {\bibinfo {volume} {7}},\ \bibinfo {pages} {201138} (\bibinfo {year}
  {2020})}\BibitemShut {NoStop}%
\bibitem [{\citenamefont {Erd{\"{o}}s}\ and\ \citenamefont
  {R{\'{e}}nyi}(1959)}]{Erdos1959OnI}%
  \BibitemOpen
  \bibfield  {author} {\bibinfo {author} {\bibfnamefont {P.}~\bibnamefont
  {Erd{\"{o}}s}}\ and\ \bibinfo {author} {\bibfnamefont {A.}~\bibnamefont
  {R{\'{e}}nyi}},\ }\bibfield  {title} {\bibinfo {title} {{On random graphs
  I}},\ }\href@noop {} {\bibfield  {journal} {\bibinfo  {journal}
  {Publicationes Mathematicae}\ }\textbf {\bibinfo {volume} {6}} (\bibinfo
  {year} {1959})}\BibitemShut {NoStop}%
\bibitem [{\citenamefont {Johnson}\ and\ \citenamefont
  {Jones}(2017)}]{Johnson2017LooplessnessCoherence}%
  \BibitemOpen
  \bibfield  {author} {\bibinfo {author} {\bibfnamefont {S.}~\bibnamefont
  {Johnson}}\ and\ \bibinfo {author} {\bibfnamefont {N.~S.}\ \bibnamefont
  {Jones}},\ }\bibfield  {title} {\bibinfo {title} {{Looplessness in networks
  is linked to trophic coherence}},\ }\bibfield  {journal} {\bibinfo  {journal}
  {Proceedings of the National Academy of Sciences of the United States of
  America}\ }\textbf {\bibinfo {volume} {114}},\ \href
  {https://doi.org/10.1073/pnas.1613786114} {10.1073/pnas.1613786114} (\bibinfo
  {year} {2017})\BibitemShut {NoStop}%
\bibitem [{\citenamefont
  {Johnson}(2020{\natexlab{a}})}]{Johnson2020DigraphsSystems}%
  \BibitemOpen
  \bibfield  {author} {\bibinfo {author} {\bibfnamefont {S.}~\bibnamefont
  {Johnson}},\ }\bibfield  {title} {\bibinfo {title} {{Digraphs are different:
  why directionality matters in complex systems}},\ }\href
  {https://doi.org/10.1088/2632-072x/ab8e2f} {\bibfield  {journal} {\bibinfo
  {journal} {Journal of Physics: Complexity}\ }\textbf {\bibinfo {volume}
  {1}},\ \bibinfo {pages} {015003} (\bibinfo {year}
  {2020}{\natexlab{a}})}\BibitemShut {NoStop}%
\bibitem [{\citenamefont {Kale}\ \emph {et~al.}(2018)\citenamefont {Kale},
  \citenamefont {Zalesky},\ and\ \citenamefont
  {Gollo}}]{Kale2018EstimatingConnectomes}%
  \BibitemOpen
  \bibfield  {author} {\bibinfo {author} {\bibfnamefont {P.}~\bibnamefont
  {Kale}}, \bibinfo {author} {\bibfnamefont {A.}~\bibnamefont {Zalesky}},\ and\
  \bibinfo {author} {\bibfnamefont {L.~L.}\ \bibnamefont {Gollo}},\ }\bibfield
  {title} {\bibinfo {title} {{Estimating the impact of structural
  directionality: How reliable are undirected connectomes?}},\ }\bibfield
  {journal} {\bibinfo  {journal} {Network Neuroscience}\ }\textbf {\bibinfo
  {volume} {2}},\ \href {https://doi.org/10.1162/netn{\_}a{\_}00040}
  {10.1162/netn{\_}a{\_}00040} (\bibinfo {year} {2018})\BibitemShut {NoStop}%
\bibitem [{\citenamefont {Hopfield}(1982)}]{Hopfield1982NeuralAbilities}%
  \BibitemOpen
  \bibfield  {author} {\bibinfo {author} {\bibfnamefont {J.~J.}\ \bibnamefont
  {Hopfield}},\ }\bibfield  {title} {\bibinfo {title} {{Neural networks and
  physical systems with emergent collective computational abilities}},\
  }\href@noop {} {\bibfield  {journal} {\bibinfo  {journal} {Proc. NatL Acad.
  Sci. USA}\ }\textbf {\bibinfo {volume} {79}},\ \bibinfo {pages} {2554}
  (\bibinfo {year} {1982})}\BibitemShut {NoStop}%
\bibitem [{\citenamefont {Zheng}\ \emph {et~al.}(2010)\citenamefont {Zheng},
  \citenamefont {Zhang},\ and\ \citenamefont
  {Tang}}]{Zheng2010AnalysisDynamics}%
  \BibitemOpen
  \bibfield  {author} {\bibinfo {author} {\bibfnamefont {P.}~\bibnamefont
  {Zheng}}, \bibinfo {author} {\bibfnamefont {J.}~\bibnamefont {Zhang}},\ and\
  \bibinfo {author} {\bibfnamefont {W.}~\bibnamefont {Tang}},\ }\bibfield
  {title} {\bibinfo {title} {{Analysis and design of asymmetric Hopfield
  networks with discrete-time dynamics}},\ }\href
  {https://doi.org/10.1007/s00422-010-0391-9} {\bibfield  {journal} {\bibinfo
  {journal} {Biol Cybern}\ }\textbf {\bibinfo {volume} {103}},\ \bibinfo
  {pages} {79} (\bibinfo {year} {2010})}\BibitemShut {NoStop}%
\bibitem [{\citenamefont {Cantini}\ and\ \citenamefont
  {Caselle}(2019)}]{Cantini2019Hope4Genes:Data}%
  \BibitemOpen
  \bibfield  {author} {\bibinfo {author} {\bibfnamefont {L.}~\bibnamefont
  {Cantini}}\ and\ \bibinfo {author} {\bibfnamefont {M.}~\bibnamefont
  {Caselle}},\ }\bibfield  {title} {\bibinfo {title} {{Hope4Genes: a
  Hopfield-like class prediction algorithm for transcriptomic data}},\
  }\bibfield  {journal} {\bibinfo  {journal} {Scientific Reports}\ }\textbf
  {\bibinfo {volume} {9}},\ \href {https://doi.org/10.1038/s41598-018-36744-y}
  {10.1038/s41598-018-36744-y} (\bibinfo {year} {2019})\BibitemShut {NoStop}%
\bibitem [{\citenamefont {Szedlak}\ \emph {et~al.}(2014)\citenamefont
  {Szedlak}, \citenamefont {Paternostro},\ and\ \citenamefont
  {Piermarocchi}}]{Szedlak2014ControlAttractors}%
  \BibitemOpen
  \bibfield  {author} {\bibinfo {author} {\bibfnamefont {A.}~\bibnamefont
  {Szedlak}}, \bibinfo {author} {\bibfnamefont {G.}~\bibnamefont
  {Paternostro}},\ and\ \bibinfo {author} {\bibfnamefont {C.}~\bibnamefont
  {Piermarocchi}},\ }\bibfield  {title} {\bibinfo {title} {{Control of
  Asymmetric Hopfield Networks and Application to Cancer Attractors}},\ }\href
  {https://doi.org/10.1371/journal.pone.0105842} {\bibfield  {journal}
  {\bibinfo  {journal} {PLoS ONE}\ }\textbf {\bibinfo {volume} {9}},\ \bibinfo
  {pages} {105842} (\bibinfo {year} {2014})}\BibitemShut {NoStop}%
\bibitem [{\citenamefont {Baggio}\ \emph {et~al.}(2020)\citenamefont {Baggio},
  \citenamefont {Rutten}, \citenamefont {Rutten}, \citenamefont {Hennequin},\
  and\ \citenamefont {Zampieri}}]{Baggio2020EfficientNon-normality}%
  \BibitemOpen
  \bibfield  {author} {\bibinfo {author} {\bibfnamefont {G.}~\bibnamefont
  {Baggio}}, \bibinfo {author} {\bibfnamefont {V.}~\bibnamefont {Rutten}},
  \bibinfo {author} {\bibfnamefont {V.}~\bibnamefont {Rutten}}, \bibinfo
  {author} {\bibfnamefont {G.}~\bibnamefont {Hennequin}},\ and\ \bibinfo
  {author} {\bibfnamefont {S.}~\bibnamefont {Zampieri}},\ }\bibfield  {title}
  {\bibinfo {title} {{Efficient communication over complex dynamical networks:
  The role of matrix non-normality}},\ }\bibfield  {journal} {\bibinfo
  {journal} {Science Advances}\ }\textbf {\bibinfo {volume} {6}},\ \href
  {https://doi.org/10.1126/sciadv.aba2282} {10.1126/sciadv.aba2282} (\bibinfo
  {year} {2020})\BibitemShut {NoStop}%
\bibitem [{\citenamefont {Liu}\ and\ \citenamefont
  {Barab{\'{a}}si}(2016)}]{Liu2016ControlSystems}%
  \BibitemOpen
  \bibfield  {author} {\bibinfo {author} {\bibfnamefont {Y.~Y.}\ \bibnamefont
  {Liu}}\ and\ \bibinfo {author} {\bibfnamefont {A.~L.}\ \bibnamefont
  {Barab{\'{a}}si}},\ }\bibfield  {title} {\bibinfo {title} {{Control
  principles of complex systems}},\ }\href
  {https://doi.org/10.1103/RevModPhys.88.035006} {\bibfield  {journal}
  {\bibinfo  {journal} {Reviews of Modern Physics}\ }\textbf {\bibinfo {volume}
  {88}},\ \bibinfo {pages} {035006} (\bibinfo {year} {2016})}\BibitemShut
  {NoStop}%
\bibitem [{\citenamefont {Lynn}\ and\ \citenamefont
  {Bassett}(2019)}]{Lynn2019TheControl}%
  \BibitemOpen
  \bibfield  {author} {\bibinfo {author} {\bibfnamefont {C.~W.}\ \bibnamefont
  {Lynn}}\ and\ \bibinfo {author} {\bibfnamefont {D.~S.}\ \bibnamefont
  {Bassett}},\ }\bibfield  {title} {\bibinfo {title} {{The physics of brain
  network structure, function and control}},\ }\bibfield  {journal} {\bibinfo
  {journal} {Nature Reviews Physics}\ }\textbf {\bibinfo {volume} {1}},\ \href
  {https://doi.org/10.1038/s42254-019-0040-8} {10.1038/s42254-019-0040-8}
  (\bibinfo {year} {2019})\BibitemShut {NoStop}%
\bibitem [{\citenamefont {Leonetti}\ \emph {et~al.}(2020)\citenamefont
  {Leonetti}, \citenamefont {Folli}, \citenamefont {Milanetti}, \citenamefont
  {Ruocco},\ and\ \citenamefont {Gosti}}]{Leonetti2020NetworkBrain}%
  \BibitemOpen
  \bibfield  {author} {\bibinfo {author} {\bibfnamefont {M.}~\bibnamefont
  {Leonetti}}, \bibinfo {author} {\bibfnamefont {V.}~\bibnamefont {Folli}},
  \bibinfo {author} {\bibfnamefont {E.}~\bibnamefont {Milanetti}}, \bibinfo
  {author} {\bibfnamefont {G.}~\bibnamefont {Ruocco}},\ and\ \bibinfo {author}
  {\bibfnamefont {G.}~\bibnamefont {Gosti}},\ }\bibfield  {title} {\bibinfo
  {title} {{Network dilution and asymmetry in an efficient brain}},\ }\bibfield
   {journal} {\bibinfo  {journal} {Philosophical Magazine}\ }\textbf {\bibinfo
  {volume} {100}},\ \href {https://doi.org/10.1080/14786435.2020.1750726}
  {10.1080/14786435.2020.1750726} (\bibinfo {year} {2020})\BibitemShut
  {NoStop}%
\bibitem [{\citenamefont {Klaise}\ and\ \citenamefont
  {Johnson}(2016)}]{Klaise2016FromProcesses}%
  \BibitemOpen
  \bibfield  {author} {\bibinfo {author} {\bibfnamefont {J.}~\bibnamefont
  {Klaise}}\ and\ \bibinfo {author} {\bibfnamefont {S.}~\bibnamefont
  {Johnson}},\ }\bibfield  {title} {\bibinfo {title} {{From neurons to
  epidemics: How trophic coherence affects spreading processes}},\ }\bibfield
  {journal} {\bibinfo  {journal} {Chaos}\ }\textbf {\bibinfo {volume} {26}},\
  \href {https://doi.org/10.1063/1.4953160} {10.1063/1.4953160} (\bibinfo
  {year} {2016})\BibitemShut {NoStop}%
\bibitem [{\citenamefont {Chau}\ and\ \citenamefont
  {Chen}(2007)}]{Chau2007IncorporatingSearching}%
  \BibitemOpen
  \bibfield  {author} {\bibinfo {author} {\bibfnamefont {M.}~\bibnamefont
  {Chau}}\ and\ \bibinfo {author} {\bibfnamefont {H.}~\bibnamefont {Chen}},\
  }\bibfield  {title} {\bibinfo {title} {{Incorporating web analysis into
  neural networks: An example in hopfield net searching}},\ }\bibfield
  {journal} {\bibinfo  {journal} {IEEE Transactions on Systems, Man and
  Cybernetics Part C: Applications and Reviews}\ }\textbf {\bibinfo {volume}
  {37}},\ \href {https://doi.org/10.1109/TSMCC.2007.893277}
  {10.1109/TSMCC.2007.893277} (\bibinfo {year} {2007})\BibitemShut {NoStop}%
\bibitem [{\citenamefont {Szedlak}\ \emph {et~al.}(2017)\citenamefont
  {Szedlak}, \citenamefont {Sims}, \citenamefont {Smith}, \citenamefont
  {Paternostro},\ and\ \citenamefont {Piermarocchi}}]{Szedlak2017CellSystems}%
  \BibitemOpen
  \bibfield  {author} {\bibinfo {author} {\bibfnamefont {A.}~\bibnamefont
  {Szedlak}}, \bibinfo {author} {\bibfnamefont {S.}~\bibnamefont {Sims}},
  \bibinfo {author} {\bibfnamefont {N.}~\bibnamefont {Smith}}, \bibinfo
  {author} {\bibfnamefont {G.}~\bibnamefont {Paternostro}},\ and\ \bibinfo
  {author} {\bibfnamefont {C.}~\bibnamefont {Piermarocchi}},\ }\bibfield
  {title} {\bibinfo {title} {{Cell cycle time series gene expression data
  encoded as cyclic attractors in Hopfield systems}},\ }\bibfield  {journal}
  {\bibinfo  {journal} {PLoS Computational Biology}\ }\textbf {\bibinfo
  {volume} {13}},\ \href {https://doi.org/10.1371/journal.pcbi.1005849}
  {10.1371/journal.pcbi.1005849} (\bibinfo {year} {2017})\BibitemShut {NoStop}%
\bibitem [{\citenamefont {Rockne}\ \emph {et~al.}(2020)\citenamefont {Rockne},
  \citenamefont {Ao}, \citenamefont {Espinosa-Soto}, \citenamefont {Alves
  Barbosa Da~Silva}, \citenamefont {Conforte}, \citenamefont {Alves},
  \citenamefont {Coelho},\ and\ \citenamefont
  {Carels}}]{Rockne2020ModelingNetworks}%
  \BibitemOpen
  \bibfield  {author} {\bibinfo {author} {\bibfnamefont {R.~C.}\ \bibnamefont
  {Rockne}}, \bibinfo {author} {\bibfnamefont {P.}~\bibnamefont {Ao}}, \bibinfo
  {author} {\bibfnamefont {C.}~\bibnamefont {Espinosa-Soto}}, \bibinfo {author}
  {\bibfnamefont {F.}~\bibnamefont {Alves Barbosa Da~Silva}}, \bibinfo {author}
  {\bibfnamefont {A.~J.}\ \bibnamefont {Conforte}}, \bibinfo {author}
  {\bibfnamefont {L.}~\bibnamefont {Alves}}, \bibinfo {author} {\bibfnamefont
  {F.~C.}\ \bibnamefont {Coelho}},\ and\ \bibinfo {author} {\bibfnamefont
  {N.}~\bibnamefont {Carels}},\ }\bibfield  {title} {\bibinfo {title}
  {{Modeling Basins of Attraction for Breast Cancer Using Hopfield Networks}},\
  }\href {https://doi.org/10.3389/fgene.2020.00314} {\bibfield  {journal}
  {\bibinfo  {journal} {Frontiers in Genetics | www.frontiersin.org}\ }\textbf
  {\bibinfo {volume} {1}},\ \bibinfo {pages} {314} (\bibinfo {year}
  {2020})}\BibitemShut {NoStop}%
\bibitem [{\citenamefont {Johnson}\ \emph {et~al.}(2014)\citenamefont
  {Johnson}, \citenamefont {Dom{\'{i}}nguez-Garc{\'{i}}a}, \citenamefont
  {Donetti},\ and\ \citenamefont {Mu{\~{n}}oz}}]{Johnson2014TrophicStability}%
  \BibitemOpen
  \bibfield  {author} {\bibinfo {author} {\bibfnamefont {S.}~\bibnamefont
  {Johnson}}, \bibinfo {author} {\bibfnamefont {V.}~\bibnamefont
  {Dom{\'{i}}nguez-Garc{\'{i}}a}}, \bibinfo {author} {\bibfnamefont
  {L.}~\bibnamefont {Donetti}},\ and\ \bibinfo {author} {\bibfnamefont {M.~A.}\
  \bibnamefont {Mu{\~{n}}oz}},\ }\bibfield  {title} {\bibinfo {title} {{Trophic
  coherence determines food-web stability}},\ }\href
  {https://doi.org/10.1073/pnas.1409077111} {\bibfield  {journal} {\bibinfo
  {journal} {Proceedings of the National Academy of Sciences of the United
  States of America}\ }\textbf {\bibinfo {volume} {111}},\ \bibinfo {pages}
  {17923} (\bibinfo {year} {2014})}\BibitemShut {NoStop}%
\bibitem [{\citenamefont {Levine}(1980)}]{levine1980several}%
  \BibitemOpen
  \bibfield  {author} {\bibinfo {author} {\bibfnamefont {S.}~\bibnamefont
  {Levine}},\ }\bibfield  {title} {\bibinfo {title} {Several measures of
  trophic structure applicable to complex food webs},\ }\href@noop {}
  {\bibfield  {journal} {\bibinfo  {journal} {Journal of theoretical Biology}\
  }\textbf {\bibinfo {volume} {83}},\ \bibinfo {pages} {195} (\bibinfo {year}
  {1980})}\BibitemShut {NoStop}%
\bibitem [{\citenamefont {Pagani}\ \emph {et~al.}(2019)\citenamefont {Pagani},
  \citenamefont {Mosquera}, \citenamefont {Alturki}, \citenamefont {Johnson},
  \citenamefont {Jarvis}, \citenamefont {Wilson}, \citenamefont {Guo},\ and\
  \citenamefont {Varga}}]{Pagani2019ResilienceNetworks}%
  \BibitemOpen
  \bibfield  {author} {\bibinfo {author} {\bibfnamefont {A.}~\bibnamefont
  {Pagani}}, \bibinfo {author} {\bibfnamefont {G.}~\bibnamefont {Mosquera}},
  \bibinfo {author} {\bibfnamefont {A.}~\bibnamefont {Alturki}}, \bibinfo
  {author} {\bibfnamefont {S.}~\bibnamefont {Johnson}}, \bibinfo {author}
  {\bibfnamefont {S.}~\bibnamefont {Jarvis}}, \bibinfo {author} {\bibfnamefont
  {A.}~\bibnamefont {Wilson}}, \bibinfo {author} {\bibfnamefont
  {W.}~\bibnamefont {Guo}},\ and\ \bibinfo {author} {\bibfnamefont
  {L.}~\bibnamefont {Varga}},\ }\bibfield  {title} {\bibinfo {title}
  {{Resilience or robustness: Identifying topological vulnerabilities in rail
  networks}},\ }\bibfield  {journal} {\bibinfo  {journal} {Royal Society Open
  Science}\ }\textbf {\bibinfo {volume} {6}},\ \href
  {https://doi.org/10.1098/rsos.181301} {10.1098/rsos.181301} (\bibinfo {year}
  {2019})\BibitemShut {NoStop}%
\bibitem [{\citenamefont {Pagani}\ \emph {et~al.}(2020)\citenamefont {Pagani},
  \citenamefont {Meng}, \citenamefont {Fu}, \citenamefont {Musolesi},\ and\
  \citenamefont {Guo}}]{Pagani2020QuantifyingNetworks}%
  \BibitemOpen
  \bibfield  {author} {\bibinfo {author} {\bibfnamefont {A.}~\bibnamefont
  {Pagani}}, \bibinfo {author} {\bibfnamefont {F.}~\bibnamefont {Meng}},
  \bibinfo {author} {\bibfnamefont {G.}~\bibnamefont {Fu}}, \bibinfo {author}
  {\bibfnamefont {M.}~\bibnamefont {Musolesi}},\ and\ \bibinfo {author}
  {\bibfnamefont {W.}~\bibnamefont {Guo}},\ }\bibfield  {title} {\bibinfo
  {title} {{Quantifying Resilience via Multiscale Feedback Loops in Water
  Distribution Networks}},\ }\bibfield  {journal} {\bibinfo  {journal} {Journal
  of Water Resources Planning and Management}\ }\textbf {\bibinfo {volume}
  {146}},\ \href {https://doi.org/10.1061/(ASCE)WR.1943-5452.0001231}
  {10.1061/(ASCE)WR.1943-5452.0001231} (\bibinfo {year} {2020})\BibitemShut
  {NoStop}%
\bibitem [{\citenamefont {Pilgrim}\ \emph {et~al.}(2020)\citenamefont
  {Pilgrim}, \citenamefont {Guo},\ and\ \citenamefont
  {Johnson}}]{Pilgrim2020OrganisationalAnarchy}%
  \BibitemOpen
  \bibfield  {author} {\bibinfo {author} {\bibfnamefont {C.}~\bibnamefont
  {Pilgrim}}, \bibinfo {author} {\bibfnamefont {W.}~\bibnamefont {Guo}},\ and\
  \bibinfo {author} {\bibfnamefont {S.}~\bibnamefont {Johnson}},\ }\bibfield
  {title} {\bibinfo {title} {{Organisational Social Influence on Directed
  Hierarchical Graphs, from Tyranny to Anarchy}},\ }\bibfield  {journal}
  {\bibinfo  {journal} {Scientific Reports}\ }\textbf {\bibinfo {volume}
  {10}},\ \href {https://doi.org/10.1038/s41598-020-61196-8}
  {10.1038/s41598-020-61196-8} (\bibinfo {year} {2020})\BibitemShut {NoStop}%
\bibitem [{\citenamefont {Amit}\ \emph {et~al.}(1985)\citenamefont {Amit},
  \citenamefont {Gutfreund},\ and\ \citenamefont
  {Sompolinsky}}]{Amit1985Spin-glassNetworks}%
  \BibitemOpen
  \bibfield  {author} {\bibinfo {author} {\bibfnamefont {D.~J.}\ \bibnamefont
  {Amit}}, \bibinfo {author} {\bibfnamefont {H.}~\bibnamefont {Gutfreund}},\
  and\ \bibinfo {author} {\bibfnamefont {H.}~\bibnamefont {Sompolinsky}},\
  }\bibfield  {title} {\bibinfo {title} {{Spin-glass models of neural
  networks}},\ }\href {https://doi.org/10.1103/PhysRevA.32.1007} {\bibfield
  {journal} {\bibinfo  {journal} {Physical Review A}\ }\textbf {\bibinfo
  {volume} {32}},\ \bibinfo {pages} {1007} (\bibinfo {year}
  {1985})}\BibitemShut {NoStop}%
\bibitem [{\citenamefont {Hastings}(1970)}]{Hastings1970MonteApplications}%
  \BibitemOpen
  \bibfield  {author} {\bibinfo {author} {\bibfnamefont {W.~K.}\ \bibnamefont
  {Hastings}},\ }\bibfield  {title} {\bibinfo {title} {{Monte carlo sampling
  methods using Markov chains and their applications}},\ }\bibfield  {journal}
  {\bibinfo  {journal} {Biometrika}\ }\textbf {\bibinfo {volume} {57}},\ \href
  {https://doi.org/10.1093/biomet/57.1.97} {10.1093/biomet/57.1.97} (\bibinfo
  {year} {1970})\BibitemShut {NoStop}%
\bibitem [{\citenamefont {Ackley}\ \emph {et~al.}(1985)\citenamefont {Ackley},
  \citenamefont {Hinton},\ and\ \citenamefont
  {Sejnowski}}]{Ackley1985AMachines}%
  \BibitemOpen
  \bibfield  {author} {\bibinfo {author} {\bibfnamefont {D.~H.}\ \bibnamefont
  {Ackley}}, \bibinfo {author} {\bibfnamefont {G.~E.}\ \bibnamefont {Hinton}},\
  and\ \bibinfo {author} {\bibfnamefont {T.~J.}\ \bibnamefont {Sejnowski}},\
  }\bibfield  {title} {\bibinfo {title} {{A learning algorithm for boltzmann
  machines}},\ }\href {https://doi.org/10.1016/S0364-0213(85)80012-4}
  {\bibfield  {journal} {\bibinfo  {journal} {Cognitive Science}\ }\textbf
  {\bibinfo {volume} {9}},\ \bibinfo {pages} {147} (\bibinfo {year}
  {1985})}\BibitemShut {NoStop}%
\bibitem [{\citenamefont {Crisanti}\ \emph {et~al.}(1993)\citenamefont
  {Crisanti}, \citenamefont {Falcioni},\ and\ \citenamefont
  {Vulpiani}}]{Crisanti1993TransitionModel}%
  \BibitemOpen
  \bibfield  {author} {\bibinfo {author} {\bibfnamefont {A.}~\bibnamefont
  {Crisanti}}, \bibinfo {author} {\bibfnamefont {M.}~\bibnamefont {Falcioni}},\
  and\ \bibinfo {author} {\bibfnamefont {A.}~\bibnamefont {Vulpiani}},\
  }\bibfield  {title} {\bibinfo {title} {{Transition from regular to complex
  behaviour in a discrete deterministic asymmetric neural network model}},\
  }\href {https://doi.org/10.1088/0305-4470/26/14/011} {\bibfield  {journal}
  {\bibinfo  {journal} {Journal of Physics A: Mathematical and General}\
  }\textbf {\bibinfo {volume} {26}},\ \bibinfo {pages} {3441} (\bibinfo {year}
  {1993})}\BibitemShut {NoStop}%
\bibitem [{\citenamefont {Attneave}\ \emph {et~al.}(1950)\citenamefont
  {Attneave}, \citenamefont {B.},\ and\ \citenamefont
  {Hebb}}]{Attneave1950TheTheory}%
  \BibitemOpen
  \bibfield  {author} {\bibinfo {author} {\bibfnamefont {F.}~\bibnamefont
  {Attneave}}, \bibinfo {author} {\bibfnamefont {M.}~\bibnamefont {B.}},\ and\
  \bibinfo {author} {\bibfnamefont {D.~O.}\ \bibnamefont {Hebb}},\ }\bibfield
  {title} {\bibinfo {title} {{The Organization of Behavior; A
  Neuropsychological Theory}},\ }\bibfield  {journal} {\bibinfo  {journal} {The
  American Journal of Psychology}\ }\textbf {\bibinfo {volume} {63}},\ \href
  {https://doi.org/10.2307/1418888} {10.2307/1418888} (\bibinfo {year}
  {1950})\BibitemShut {NoStop}%
\bibitem [{\citenamefont {Tanaka}\ \emph {et~al.}(2020)\citenamefont {Tanaka},
  \citenamefont {Nakane}, \citenamefont {Takeuchi}, \citenamefont {Yamane},
  \citenamefont {Nakano}, \citenamefont {Katayama},\ and\ \citenamefont
  {Hirose}}]{Tanaka2020SpatiallyMemory}%
  \BibitemOpen
  \bibfield  {author} {\bibinfo {author} {\bibfnamefont {G.}~\bibnamefont
  {Tanaka}}, \bibinfo {author} {\bibfnamefont {R.}~\bibnamefont {Nakane}},
  \bibinfo {author} {\bibfnamefont {T.}~\bibnamefont {Takeuchi}}, \bibinfo
  {author} {\bibfnamefont {T.}~\bibnamefont {Yamane}}, \bibinfo {author}
  {\bibfnamefont {D.}~\bibnamefont {Nakano}}, \bibinfo {author} {\bibfnamefont
  {Y.}~\bibnamefont {Katayama}},\ and\ \bibinfo {author} {\bibfnamefont
  {A.}~\bibnamefont {Hirose}},\ }\bibfield  {title} {\bibinfo {title}
  {{Spatially Arranged Sparse Recurrent Neural Networks for Energy Efficient
  Associative Memory}},\ }\bibfield  {journal} {\bibinfo  {journal} {IEEE
  Transactions on Neural Networks and Learning Systems}\ }\textbf {\bibinfo
  {volume} {31}},\ \href {https://doi.org/10.1109/TNNLS.2019.2899344}
  {10.1109/TNNLS.2019.2899344} (\bibinfo {year} {2020})\BibitemShut {NoStop}%
\bibitem [{\citenamefont {Diederich}\ and\ \citenamefont
  {Opper}(1987)}]{Diederich1987LearningRules}%
  \BibitemOpen
  \bibfield  {author} {\bibinfo {author} {\bibfnamefont {S.}~\bibnamefont
  {Diederich}}\ and\ \bibinfo {author} {\bibfnamefont {M.}~\bibnamefont
  {Opper}},\ }\bibfield  {title} {\bibinfo {title} {{Learning of correlated
  patterns in spin-glass networks by local learning rules}},\ }\href
  {https://doi.org/10.1103/PhysRevLett.58.949} {\bibfield  {journal} {\bibinfo
  {journal} {Physical Review Letters}\ }\textbf {\bibinfo {volume} {58}},\
  \bibinfo {pages} {949} (\bibinfo {year} {1987})}\BibitemShut {NoStop}%
\bibitem [{\citenamefont {Gardner}(1988)}]{Gardner1988TheModels}%
  \BibitemOpen
  \bibfield  {author} {\bibinfo {author} {\bibfnamefont {E.}~\bibnamefont
  {Gardner}},\ }\bibfield  {title} {\bibinfo {title} {{The space of
  interactions in neural network models}},\ }\bibfield  {journal} {\bibinfo
  {journal} {Journal of Physics A: General Physics}\ }\textbf {\bibinfo
  {volume} {21}},\ \href {https://doi.org/10.1088/0305-4470/21/1/030}
  {10.1088/0305-4470/21/1/030} (\bibinfo {year} {1988})\BibitemShut {NoStop}%
\bibitem [{\citenamefont {Davey}\ and\ \citenamefont
  {Adams}(2004)}]{Davey2004HighConstraints}%
  \BibitemOpen
  \bibfield  {author} {\bibinfo {author} {\bibfnamefont {N.}~\bibnamefont
  {Davey}}\ and\ \bibinfo {author} {\bibfnamefont {R.}~\bibnamefont {Adams}},\
  }\bibfield  {title} {\bibinfo {title} {{High capacity associative memories
  and connection constraints}},\ }\bibfield  {journal} {\bibinfo  {journal}
  {Connection Science}\ }\textbf {\bibinfo {volume} {16}},\ \href
  {https://doi.org/10.1080/09540090310001659981} {10.1080/09540090310001659981}
  (\bibinfo {year} {2004})\BibitemShut {NoStop}%
\bibitem [{\citenamefont {Markram}\ \emph {et~al.}(1997)\citenamefont
  {Markram}, \citenamefont {L{\"u}bke}, \citenamefont {Frotscher},\ and\
  \citenamefont {Sakmann}}]{markram1997regulation}%
  \BibitemOpen
  \bibfield  {author} {\bibinfo {author} {\bibfnamefont {H.}~\bibnamefont
  {Markram}}, \bibinfo {author} {\bibfnamefont {J.}~\bibnamefont {L{\"u}bke}},
  \bibinfo {author} {\bibfnamefont {M.}~\bibnamefont {Frotscher}},\ and\
  \bibinfo {author} {\bibfnamefont {B.}~\bibnamefont {Sakmann}},\ }\bibfield
  {title} {\bibinfo {title} {Regulation of synaptic efficacy by coincidence of
  postsynaptic aps and epsps},\ }\href@noop {} {\bibfield  {journal} {\bibinfo
  {journal} {Science}\ }\textbf {\bibinfo {volume} {275}},\ \bibinfo {pages}
  {213} (\bibinfo {year} {1997})}\BibitemShut {NoStop}%
\bibitem [{\citenamefont {Gutfreund}(1988)}]{Gutfreund1988NeuralPatterns}%
  \BibitemOpen
  \bibfield  {author} {\bibinfo {author} {\bibfnamefont {H.}~\bibnamefont
  {Gutfreund}},\ }\bibfield  {title} {\bibinfo {title} {{Neural networks with
  hierarchically correlated patterns}},\ }\href
  {https://doi.org/10.1103/PhysRevA.37.570} {\bibfield  {journal} {\bibinfo
  {journal} {Physical Review A}\ }\textbf {\bibinfo {volume} {37}},\ \bibinfo
  {pages} {570} (\bibinfo {year} {1988})}\BibitemShut {NoStop}%
\bibitem [{\citenamefont {Klaise}\ and\ \citenamefont
  {Johnson}(2017)}]{Klaise2017TheWebs}%
  \BibitemOpen
  \bibfield  {author} {\bibinfo {author} {\bibfnamefont {J.}~\bibnamefont
  {Klaise}}\ and\ \bibinfo {author} {\bibfnamefont {S.}~\bibnamefont
  {Johnson}},\ }\bibfield  {title} {\bibinfo {title} {{The origin of motif
  families in food webs}},\ }\bibfield  {journal} {\bibinfo  {journal}
  {Scientific Reports}\ }\textbf {\bibinfo {volume} {7}},\ \href
  {https://doi.org/10.1038/s41598-017-15496-1} {10.1038/s41598-017-15496-1}
  (\bibinfo {year} {2017})\BibitemShut {NoStop}%
\bibitem [{\citenamefont {Wright}\ \emph {et~al.}(2019)\citenamefont {Wright},
  \citenamefont {Yoon}, \citenamefont {Ferreira}, \citenamefont {Mendes},\ and\
  \citenamefont {Goltsev}}]{Wright2019TheDynamics}%
  \BibitemOpen
  \bibfield  {author} {\bibinfo {author} {\bibfnamefont {E.~A.}\ \bibnamefont
  {Wright}}, \bibinfo {author} {\bibfnamefont {S.}~\bibnamefont {Yoon}},
  \bibinfo {author} {\bibfnamefont {A.~L.}\ \bibnamefont {Ferreira}}, \bibinfo
  {author} {\bibfnamefont {J.~F.}\ \bibnamefont {Mendes}},\ and\ \bibinfo
  {author} {\bibfnamefont {A.~V.}\ \bibnamefont {Goltsev}},\ }\bibfield
  {title} {\bibinfo {title} {{The central role of peripheral nodes in directed
  network dynamics}},\ }\bibfield  {journal} {\bibinfo  {journal} {Scientific
  Reports}\ }\textbf {\bibinfo {volume} {9}},\ \href
  {https://doi.org/10.1038/s41598-019-49537-8} {10.1038/s41598-019-49537-8}
  (\bibinfo {year} {2019})\BibitemShut {NoStop}%
\bibitem [{\citenamefont {P{\'{e}}rez~Castillo}\ and\ \citenamefont
  {Skantzos}(2004)}]{PerezCastillo2004TheGraph}%
  \BibitemOpen
  \bibfield  {author} {\bibinfo {author} {\bibfnamefont {I.}~\bibnamefont
  {P{\'{e}}rez~Castillo}}\ and\ \bibinfo {author} {\bibfnamefont {N.~S.}\
  \bibnamefont {Skantzos}},\ }\bibfield  {title} {\bibinfo {title} {{The
  Little-Hopfield model on a sparse random graph}},\ }\bibfield  {journal}
  {\bibinfo  {journal} {Journal of Physics A: Mathematical and General}\
  }\textbf {\bibinfo {volume} {37}},\ \href
  {https://doi.org/10.1088/0305-4470/37/39/003} {10.1088/0305-4470/37/39/003}
  (\bibinfo {year} {2004})\BibitemShut {NoStop}%
\bibitem [{\citenamefont {Mceliece}\ \emph {et~al.}(1987)\citenamefont
  {Mceliece}, \citenamefont {Posner}, \citenamefont {Rodemich},\ and\
  \citenamefont {Venkatesh}}]{Mceliece1987TheMemory}%
  \BibitemOpen
  \bibfield  {author} {\bibinfo {author} {\bibfnamefont {R.~J.}\ \bibnamefont
  {Mceliece}}, \bibinfo {author} {\bibfnamefont {E.~C.}\ \bibnamefont
  {Posner}}, \bibinfo {author} {\bibfnamefont {E.~R.}\ \bibnamefont
  {Rodemich}},\ and\ \bibinfo {author} {\bibfnamefont {S.~S.}\ \bibnamefont
  {Venkatesh}},\ }\bibfield  {title} {\bibinfo {title} {{The Capacity of the
  Hopfield Associative Memory}},\ }\bibfield  {journal} {\bibinfo  {journal}
  {IEEE Transactions on Information Theory}\ }\textbf {\bibinfo {volume}
  {33}},\ \href {https://doi.org/10.1109/TIT.1987.1057328}
  {10.1109/TIT.1987.1057328} (\bibinfo {year} {1987})\BibitemShut {NoStop}%
\bibitem [{\citenamefont {Sompolinsky}(1986)}]{Sompolinsky1986NeuralNoise}%
  \BibitemOpen
  \bibfield  {author} {\bibinfo {author} {\bibfnamefont {H.}~\bibnamefont
  {Sompolinsky}},\ }\bibfield  {title} {\bibinfo {title} {{Neural networks with
  nonlinear synapses and a static noise}},\ }\bibfield  {journal} {\bibinfo
  {journal} {Physical Review A}\ }\textbf {\bibinfo {volume} {34}},\ \href
  {https://doi.org/10.1103/PhysRevA.34.2571} {10.1103/PhysRevA.34.2571}
  (\bibinfo {year} {1986})\BibitemShut {NoStop}%
\bibitem [{\citenamefont {Asllani}\ \emph {et~al.}(2018)\citenamefont
  {Asllani}, \citenamefont {Lambiotte},\ and\ \citenamefont
  {Carletti}}]{Asllani2018StructureNetworks}%
  \BibitemOpen
  \bibfield  {author} {\bibinfo {author} {\bibfnamefont {M.}~\bibnamefont
  {Asllani}}, \bibinfo {author} {\bibfnamefont {R.}~\bibnamefont {Lambiotte}},\
  and\ \bibinfo {author} {\bibfnamefont {T.}~\bibnamefont {Carletti}},\
  }\bibfield  {title} {\bibinfo {title} {{Structure and dynamical behavior of
  non-normal networks}},\ }\bibfield  {journal} {\bibinfo  {journal} {Science
  Advances}\ }\textbf {\bibinfo {volume} {4}},\ \href
  {https://doi.org/10.1126/sciadv.aau9403} {10.1126/sciadv.aau9403} (\bibinfo
  {year} {2018})\BibitemShut {NoStop}%
\bibitem [{\citenamefont {Asllani}\ and\ \citenamefont
  {Carletti}(2018)}]{Asllani2018TopologicalSystems}%
  \BibitemOpen
  \bibfield  {author} {\bibinfo {author} {\bibfnamefont {M.}~\bibnamefont
  {Asllani}}\ and\ \bibinfo {author} {\bibfnamefont {T.}~\bibnamefont
  {Carletti}},\ }\bibfield  {title} {\bibinfo {title} {{Topological resilience
  in non-normal networked systems}},\ }\href
  {https://doi.org/10.1103/PhysRevE.97.042302} {\bibfield  {journal} {\bibinfo
  {journal} {Physical Review E}\ }\textbf {\bibinfo {volume} {97}},\ \bibinfo
  {pages} {042302} (\bibinfo {year} {2018})}\BibitemShut {NoStop}%
\bibitem [{\citenamefont {Muolo}\ \emph {et~al.}(2019)\citenamefont {Muolo},
  \citenamefont {Asllani}, \citenamefont {Fanelli}, \citenamefont {Maini},\
  and\ \citenamefont {Carletti}}]{Muolo2019PatternsSystems}%
  \BibitemOpen
  \bibfield  {author} {\bibinfo {author} {\bibfnamefont {R.}~\bibnamefont
  {Muolo}}, \bibinfo {author} {\bibfnamefont {M.}~\bibnamefont {Asllani}},
  \bibinfo {author} {\bibfnamefont {D.}~\bibnamefont {Fanelli}}, \bibinfo
  {author} {\bibfnamefont {P.~K.}\ \bibnamefont {Maini}},\ and\ \bibinfo
  {author} {\bibfnamefont {T.}~\bibnamefont {Carletti}},\ }\bibfield  {title}
  {\bibinfo {title} {{Patterns of non-normality in networked systems}},\
  }\bibfield  {journal} {\bibinfo  {journal} {Journal of Theoretical Biology}\
  }\textbf {\bibinfo {volume} {480}},\ \href
  {https://doi.org/10.1016/j.jtbi.2019.07.004} {10.1016/j.jtbi.2019.07.004}
  (\bibinfo {year} {2019})\BibitemShut {NoStop}%
\bibitem [{\citenamefont {Muolo}\ \emph {et~al.}(2020)\citenamefont {Muolo},
  \citenamefont {Carletti}, \citenamefont {Gleeson},\ and\ \citenamefont
  {Asllani}}]{Muolo2020SynchronizationOptimality}%
  \BibitemOpen
  \bibfield  {author} {\bibinfo {author} {\bibfnamefont {R.}~\bibnamefont
  {Muolo}}, \bibinfo {author} {\bibfnamefont {T.}~\bibnamefont {Carletti}},
  \bibinfo {author} {\bibfnamefont {J.~P.}\ \bibnamefont {Gleeson}},\ and\
  \bibinfo {author} {\bibfnamefont {M.}~\bibnamefont {Asllani}},\ }\bibfield
  {title} {\bibinfo {title} {{Synchronization dynamics in non-normal networks:
  The trade-off for optimality}},\ }\bibfield  {journal} {\bibinfo  {journal}
  {Entropy}\ }\textbf {\bibinfo {volume} {23}},\ \href
  {https://doi.org/10.3390/e23010036} {10.3390/e23010036} (\bibinfo {year}
  {2020})\BibitemShut {NoStop}%
\bibitem [{\citenamefont {Fruchart}\ \emph {et~al.}(2021)\citenamefont
  {Fruchart}, \citenamefont {Hanai}, \citenamefont {Littlewood},\ and\
  \citenamefont {Vitelli}}]{Fruchart2021Non-reciprocalTransitions}%
  \BibitemOpen
  \bibfield  {author} {\bibinfo {author} {\bibfnamefont {M.}~\bibnamefont
  {Fruchart}}, \bibinfo {author} {\bibfnamefont {R.}~\bibnamefont {Hanai}},
  \bibinfo {author} {\bibfnamefont {P.~B.}\ \bibnamefont {Littlewood}},\ and\
  \bibinfo {author} {\bibfnamefont {V.}~\bibnamefont {Vitelli}},\ }\bibfield
  {title} {\bibinfo {title} {{Non-reciprocal phase transitions}},\ }\bibfield
  {journal} {\bibinfo  {journal} {Nature}\ }\textbf {\bibinfo {volume} {592}},\
  \href {https://doi.org/10.1038/s41586-021-03375-9}
  {10.1038/s41586-021-03375-9} (\bibinfo {year} {2021})\BibitemShut {NoStop}%
\bibitem [{\citenamefont {Peixoto}(2014)}]{peixoto_graph-tool_2014}%
  \BibitemOpen
  \bibfield  {author} {\bibinfo {author} {\bibfnamefont {T.~P.}\ \bibnamefont
  {Peixoto}},\ }\bibfield  {title} {\bibinfo {title} {The graph-tool python
  library},\ }\bibfield  {journal} {\bibinfo  {journal} {figshare}\ }\href
  {https://doi.org/10.6084/m9.figshare.1164194} {10.6084/m9.figshare.1164194}
  (\bibinfo {year} {2014}),\ \bibinfo {note}
  {\url{http://figshare.com/articles/graph_tool/1164194}}\BibitemShut {NoStop}%
\bibitem [{\citenamefont {Hagberg}\ \emph {et~al.}(2008)\citenamefont
  {Hagberg}, \citenamefont {Schult},\ and\ \citenamefont
  {Swart}}]{Hagberg2008ExploringNetworkX}%
  \BibitemOpen
  \bibfield  {author} {\bibinfo {author} {\bibfnamefont {A.~A.}\ \bibnamefont
  {Hagberg}}, \bibinfo {author} {\bibfnamefont {D.~A.}\ \bibnamefont
  {Schult}},\ and\ \bibinfo {author} {\bibfnamefont {P.~J.}\ \bibnamefont
  {Swart}},\ }\bibfield  {title} {\bibinfo {title} {{Exploring network
  structure, dynamics, and function using NetworkX}},\ }in\ \href@noop {}
  {\emph {\bibinfo {booktitle} {7th Python in Science Conference (SciPy
  2008)}}}\ (\bibinfo {year} {2008})\BibitemShut {NoStop}%
\bibitem [{\citenamefont {Bromberger}\ and\ \citenamefont {other
  contributors}(2017)}]{Bromberger17}%
  \BibitemOpen
  \bibfield  {author} {\bibinfo {author} {\bibfnamefont {S.}~\bibnamefont
  {Bromberger}}\ and\ \bibinfo {author} {\bibnamefont {other contributors}},\
  }\href {https://doi.org/10.5281/zenodo.889971} {\bibinfo {title}
  {Juliagraphs/lightgraphs.jl: an optimized graphs package for the julia
  programming language}} (\bibinfo {year} {2017}),\ \bibinfo {note}
  {\url{https://doi.org/10.5281/zenodo.889971}}\BibitemShut {NoStop}%
\bibitem [{\citenamefont {Behnel}\ \emph {et~al.}(2011)\citenamefont {Behnel},
  \citenamefont {Bradshaw}, \citenamefont {Citro}, \citenamefont {Dalcin},
  \citenamefont {Seljebotn},\ and\ \citenamefont
  {Smith}}]{Behnel2011Cython:Worlds}%
  \BibitemOpen
  \bibfield  {author} {\bibinfo {author} {\bibfnamefont {S.}~\bibnamefont
  {Behnel}}, \bibinfo {author} {\bibfnamefont {R.}~\bibnamefont {Bradshaw}},
  \bibinfo {author} {\bibfnamefont {C.}~\bibnamefont {Citro}}, \bibinfo
  {author} {\bibfnamefont {L.}~\bibnamefont {Dalcin}}, \bibinfo {author}
  {\bibfnamefont {D.~S.}\ \bibnamefont {Seljebotn}},\ and\ \bibinfo {author}
  {\bibfnamefont {K.}~\bibnamefont {Smith}},\ }\bibfield  {title} {\bibinfo
  {title} {{Cython: The best of both worlds}},\ }\bibfield  {journal} {\bibinfo
   {journal} {Computing in Science and Engineering}\ }\textbf {\bibinfo
  {volume} {13}},\ \href {https://doi.org/10.1109/MCSE.2010.118}
  {10.1109/MCSE.2010.118} (\bibinfo {year} {2011})\BibitemShut {NoStop}%
\bibitem [{\citenamefont {Johnson}(2020{\natexlab{b}})}]{DataSamJohnson}%
  \BibitemOpen
  \bibfield  {author} {\bibinfo {author} {\bibfnamefont {S.}~\bibnamefont
  {Johnson}},\ }\href@noop {} {\emph {\bibinfo {title} {www.samuel-johnson.org
  Data Repository}}} (\bibinfo {year} {(accessed October ,
  2020)}{\natexlab{b}}),\ \bibinfo {note}
  {\url{https://www.samuel-johnson.org/data}}\BibitemShut {NoStop}%
\end{thebibliography}%

\end{document}